\documentclass[sigconf]{acmart}
\usepackage{amsmath,amsfonts}
\usepackage{adjustbox}
\usepackage{algorithm}
\usepackage[noend]{algpseudocode}
\usepackage{algorithmicx}
\usepackage[toc,page]{appendix}
\usepackage{array}
\usepackage{booktabs}
\usepackage{color,colortbl}
\usepackage{enumitem}
\usepackage[utf8]{inputenc}
\usepackage{multirow}
\usepackage[final]{pdfpages}
\usepackage{placeins}
\usepackage{tikz}
\usetikzlibrary{matrix,shapes,arrows,positioning,chains, calc, automata}

\newcommand\copyrightnotice[1]{
	\begin{tikzpicture}[remember picture,overlay]
		\node[anchor=south,yshift=10pt] at (current page.south) {\fbox{\parbox{\dimexpr\textwidth-\fboxsep-\fboxrule\relax}{#1}}};
	\end{tikzpicture}
}

\newcommand\Tstrut{\rule{0pt}{2.2ex}}         
\newcommand\Bstrut{\rule[-1.1ex]{0pt}{0pt}}   

\makeatletter
\newcounter{protocol}
\newenvironment{protocol}[1][]{
	\let\c@algorithm\c@protocol
	\makeatletter
	\renewcommand{\ALG@name}{Protocol}
	\makeatother
	\begin{algorithm}[#1]
	}{\end{algorithm}
}
\makeatother

\makeatletter
\newcounter{alg}

\tikzset{every tree node/.style={align=center}}
\DeclareMathAlphabet{\pazocal}{OMS}{zplm}{m}{n}

\newcommand*{\numCompPeers}{\ensuremath{n}}
\newcommand*{\breuerSecSharingProt}{\ensuremath{\small{\textsc{KEP-Rnd-SS}}}}
\newcommand*{\breuerHomEncProt}{\ensuremath{\small{\textsc{KEP-Rnd-HE}}}}
\newcommand*{\breuerHomEncProtFootnote}{\ensuremath{\textsc{KEP-Rnd-HE}}}
\newcommand*{\crossoverKeProtocolFootnote}{\ensuremath{\textsc{Crossover-KE}}}

\newcommand*{\rootNode}{\ensuremath{\textit{root}}}
\newcommand*{\mate}{\ensuremath{\textit{mate}}}
\newcommand*{\expo}{\ensuremath{\textit{expo}}}
\newcommand*{\nontree}{\ensuremath{\textit{nonTree}}}
\newcommand*{\queue}{\ensuremath{\textit{Q}}}
\newcommand*{\compGraphMatrix}{\ensuremath{\textnormal{A}}}
\newcommand*{\nodes}{\ensuremath{\textnormal{V}}}
\newcommand*{\queueInsert}[1]{\ensuremath{\enc{\queue}\textit{.insert}(#1)}}
\newcommand*{\queueRemove}{\ensuremath{\enc{\queue}\textit{.deleteHead()}}}

\newcommand*{\zAncX}{\ensuremath{\textit{y\_anc\_x}}}

\newcommand*{\grandfather}{\ensuremath{\textit{grandfather}}}
\newcommand*{\checkOddCycleGate}{\ensuremath{\small{\textsc{Anc-Check}}\small}}
\newcommand*{\enlargeMatchingGate}{\ensuremath{\small{\textsc{Update-M}}\small}}
\newcommand*{\shuffleGate}{\ensuremath{\small{\textsc{Shuffle}}\small}}
\newcommand*{\revShuffleGate}{\ensuremath{\small{\textsc{Reverse-Shuffle}}\small}}

\newcommand*{\found}{\ensuremath{\textit{found}}}

\newcommand*{\Next}{\ensuremath{\textit{next}}}

\newcommand*{\foundPath}{\ensuremath{\textit{augPath}}}

\newcommand*{\sharedEqualityCheck}{\ensuremath{\stackrel{?}{=}}}
\newcommand*{\sharedUnequalityCheck}{\ensuremath{\stackrel{?}{\neq}}}
\newcommand*{\sharedLessThan}{\ensuremath{\stackrel{?}{<}}}

\newcommand*{\sharedMult}{\ensuremath{ \ \cdot \ }}
\newcommand*{\sharedSubtract}{\ensuremath{-}}
\newcommand*{\sharedAdd}{\ensuremath{+}}

\newcommand*{\compCheck}{\ensuremath{\small{\textsc{Comp-Check}}}}
\newcommand*{\crossoverKeProtocol}{\ensuremath{\small{\textsc{Crossover-KE}}}}

\newcommand*{\donorbloodvec}[1]{\ensuremath{B^{d}_{#1}}}
\newcommand*{\donorbloodvecNotSpec}{\ensuremath{B^{d}}}
\newcommand*{\patientbloodvec}[1]{\ensuremath{B^{p}_{#1}}}
\newcommand*{\patientbloodvecNotSpec}{\ensuremath{B^{p}}}
\newcommand*{\antigenvec}[1]{\ensuremath{A^{d}_{#1}}}
\newcommand*{\antigenvecNotSpec}{\ensuremath{A^{d}}}
\newcommand*{\antibodyvec}[1]{\ensuremath{A^{p}_{#1}}}
\newcommand*{\antibodyvecNotSpec}{\ensuremath{A^{p}}}

\definecolor{smartgray}{RGB}{61, 62, 64}
\definecolor{smartblue}{HTML}{0B6296}
\definecolor{smartred}{HTML}{C54949}
\definecolor{smartyellow}{HTML}{d0d62a}
\definecolor{smartgreen}{HTML}{61c15d}
\definecolor{smartorange}{HTML}{ff7e00}
\definecolor{smartviolet}{HTML}{d19fe8}
\definecolor{smartbrown}{HTML}{480607}
\definecolor{smartcadet}{HTML}{5F9EA0}

\newcommand*{\bigo}[1]{\ensuremath{\mathcal{O}(#1)}}

\newcommand*{\enc}[1]{\ensuremath{[ #1 ]}}

\newcommand*{\partysymbol}[1]{\ensuremath{P_{#1}}}

\AtBeginDocument{%
  \providecommand\BibTeX{{%
    \normalfont B\kern-0.5em{\scshape i\kern-0.25em b}\kern-0.8em\TeX}}}

\setcopyright{none}
\renewcommand\footnotetextcopyrightpermission[1]{}
\begin{document}
\fancyhead{}

\title{Privacy-Preserving Maximum Matching on General Graphs and its Application to Enable Privacy-Preserving Kidney Exchange\\(Extended Version)}

\author{Malte Breuer}
\email{breuer@itsec.rwth-aachen.de}
\affiliation{%
  \institution{RWTH Aachen University}
  \city{Aachen}
  \country{Germany}
}

\author{Ulrike Meyer}
\email{meyer@itsec.rwth-aachen.de}
\affiliation{%
	\institution{RWTH Aachen University}
	\city{Aachen}
	\country{Germany}
}

\author{Susanne Wetzel}
\email{swetzel@stevens.edu}
\affiliation{%
	\institution{Stevens Institute of Technology}
	\city{Hoboken}
	\state{NJ}
	\country{USA}
}

\begin{abstract}
 	To this day, there are still some countries where the exchange of kidneys between multiple incompatible patient-donor pairs is restricted by law. Typically, legal regulations in this context are put in place to prohibit coercion and manipulation in order to prevent a market for organ trade. Yet, in countries where kidney exchange is practiced, existing platforms to facilitate such exchanges generally lack sufficient privacy mechanisms. In this paper, we propose a privacy-preserving protocol for kidney exchange that not only addresses the privacy problem of existing platforms but also is geared to lead the way in overcoming legal issues in those countries where kidney exchange is still not practiced. In our approach, we use the concept of secret sharing to distribute the medical data of patients and donors among a set of computing peers in a privacy-preserving fashion. These computing peers then execute our new Secure Multi-Party Computation (SMPC) protocol among each other to determine an optimal set of kidney exchanges. As part of our new protocol, we devise a privacy-preserving solution to the maximum matching problem on general graphs. We have implemented the protocol in the SMPC benchmarking framework MP-SPDZ and provide a comprehensive performance evaluation. Furthermore, we analyze the practicality of our protocol when used in a dynamic setting (where patients and donors arrive and depart over time) based on a data set from the United Network for Organ Sharing. 
\end{abstract}



\maketitle

\copyrightnotice{\copyright\space 2022 Association for Computing Machinery. This is the author's extended version of the work. It is posted here for your personal use. Not for redistribution. The definitive version was published in the \emph{12th ACM Conference on Data and Application Security and Privacy (CODASPY'22), April 24-26, 2022, Baltimore-Washington DC Area, United States}, https://doi.org/10.1145/3508398.3511509}

\section{Introduction}\label{sec:introduction}

According to the World Health Organization kidney disease is the 10th most common cause of death worldwide~\cite{WHO_2019}. While the preferable treatment for patients with kidney disease is transplantation, the waiting lists for post-mortem kidney donation are very long, e.g., more than 90000 patients are currently on the waiting list for a kidney transplant in the US~\cite{OPTN_2021}. An alternative to post-mortem donation is to find a friend or relative who is willing to donate one of their kidneys. While many patients find such a living donor, often this donor is not compatible with the patient's medical characteristics. 

A recent development to solve this problem is kidney exchange. The idea is to consider multiple patients with incompatible living donors (also referred to as incompatible patient-donor pairs) and find exchanges among them (e.g., crossover exchanges where the donor of one patient-donor pair donates to the patient of another pair and vice versa). While there are many countries around the world where kidney exchange is already practiced, to date there are still some countries (e.g., Germany\footnote{§8 Transplantationsgesetz (German Transplantation Law)}) where kidney exchange faces legal obstacles. This is mainly due to fear of manipulation, corruption, and coercion. In those countries where kidney exchange is already practiced, it is usually organized by large centralized platforms which are responsible for the computation of the exchanges. Thus, the patient-donor pairs have to fully trust these platforms not only with their medical data but also with the correct and fair computation of the exchanges. Furthermore, such a centralized approach makes the platforms a desirable target for attackers. Compromising a single entity allows them to access the medical data of all pairs or to manipulate the computation of the exchanges which could have life-threatening consequences for the involved~patients. 

To mitigate these shortcomings, we devise a decentralized approach for kidney exchange where medical data and exchange computation are distributed among multiple parties. This ensures privacy of the data as well as protection against manipulation, corruption, and coercion for existing platforms. Also, this may lead to an adoption of kidney exchange in countries where it is still off limits today. Specifically, this work provides four main contributions: 

First, we devise the first privacy-preserving protocol for crossover kidney exchange based on secret sharing which at the same time is the first privacy-preserving protocol for kidney exchange with polynomial communication complexity. At the heart of our protocol is a privacy-preserving solution to the maximum matching problem on general graphs. To the best of our knowledge, we are the first to solve the maximum matching problem on general graphs in a privacy-preserving fashion which we consider to be of independent interest beyond the use case of kidney exchange.

Second, we implement our protocol in the state-of-the-art SMPC benchmarking framework MP-SPDZ~\cite{KellerMPSPDZ2020} and carry out a comprehensive performance analysis. The source code is published in~\cite{Breuer_Code_2022}.

Third, we compare the implementation of our protocol to the only other privacy-preserving protocol for kidney exchange~\cite{Breuer_KEprotocol_2020} known to date. To this end, we implement the protocol from~\cite{Breuer_KEprotocol_2020} in MP-SPDZ based on secret sharing and thereby significantly improve its performance. While the protocol from~\cite{Breuer_KEprotocol_2020} solves a more general problem than our newly developed protocol, we show that our protocol considerably outperforms the protocol from~\cite{Breuer_KEprotocol_2020} for the special case of crossover exchanges.

Fourth, we establish the practicality of our protocol when used as part of a dynamic kidney exchange platform where patient-donor pairs arrive and leave over time. To this end, we run simulations based on real-world data from the United Network for Organ Sharing (UNOS) which is a major kidney exchange platform in the US. We compare the performance of a kidney exchange platform using our protocol to a conventional (non-privacy-preserving) approach and measure the number of transplants for both scenarios. As kidney exchange platforms differ substantially around the world, we run simulations for a wide range of parameters reflecting many different characteristics of various kidney exchange platforms. Our simulations show that the performance difference between our privacy-preserving approach and the conventional approach is negligible for those parameters that are most likely to occur in practice.

\section{Intuition and Approach}\label{sec:overview}

\begin{figure}[t]
	\centering
\begin{adjustbox}{width=0.75\columnwidth}
\begin{tikzpicture}
	[ 
	> = stealth, 
	auto,
	node distance = 2cm, 
	semithick 
	]
	\def\opac{30}
	\def\circlerad{1.4cm}
	\def\rectheight{0.3cm}
	\def\vertdist{1.2}
	\def\horidist{5}
	\def\bend{20}
	\def\leftoffset{0.3}
	\def\rightoffset{1.3}
	\node[draw=smartblue, circle, minimum height=\circlerad, fill=smartblue!\opac](pool1) at (0,0) {$ \text{Pool}_{t} $};
	
	\node[draw=smartblue, circle, minimum height=\circlerad, fill=smartblue!\opac](pool2) at (\horidist,0) {$ \text{Pool}_{t+1} $};
	
	\node[draw=smartorange, rounded corners, fill=smartorange!\opac, minimum height=\rectheight](waiting1) at (0.5*\horidist, \vertdist) {Match offered};
	
	\node[draw=smartgray, fill=smartgray!\opac, minimum height=\rectheight](unmatched1) at (0.5*\horidist,0) {Unmatched};
	
	\node[draw=smartgreen, fill=smartgreen!\opac, rounded corners, minimum height=\rectheight](new1) at (0.5*\horidist, -\vertdist) {New patient-donor pairs};

	\node[draw=smartred, fill=smartred!\opac, rounded corners, minimum height=\rectheight, align=left, rotate=0](left1) at (0, 1.85*\vertdist) {Transplant succeeded,\\death, donor reneged, ...};
	
	\node[draw=smartred, fill=smartred!\opac, rounded corners, minimum height=\rectheight, align=left, rotate=0](left2) at (\horidist, 1.85*\vertdist) {Transplant succeeded,\\death, donor reneged, ...};
	
	\path[->] (-\leftoffset*\horidist,\vertdist) edge[dashed] (waiting1.west); 
	
	\path[->] (-\leftoffset*\horidist,\vertdist) edge[bend right=\bend] (left1.south); 
	
	\path[->] (-\leftoffset*\horidist,\vertdist) edge[bend left=\bend] (pool1.north); 
	
	\path[->] (-\leftoffset*\horidist,0) edge[dashed] (pool1.west); 
	
	\path[->] (pool1.north) edge[] (left1.south); 
	
	\path[->] (pool1.north) edge[bend left=\bend] (waiting1.west); 
	
	\path[->] (pool1.east) edge[dashed] (unmatched1.west); 
	
	\path[->] (unmatched1.east) edge[dashed] (pool2.west); 
	
	\path[->] (waiting1.east) edge[dashed]  (\rightoffset*\horidist,\vertdist); 
	
	\path[->] (waiting1.east) edge[bend right=\bend] (left2.south); 
	
	\path[->] (waiting1.east) edge[bend left=\bend] (pool2.north); 
	
	\path[->] (pool2.north) edge[] (left2.south); 
	
	\path[->] (pool2.north) edge[bend left=\bend] (\rightoffset*\horidist,\vertdist); 
	
	\path[->] (pool2.east) edge[dashed] (\rightoffset*\horidist,0); 
	
	\path[->] (new1.east) edge[bend right=\bend] (pool2.south);

\end{tikzpicture}
\end{adjustbox}
	\caption{Model of a dynamic kidney exchange platform, adapted from~\cite{Dickerson_FailureAwareKE_2019}.}
	\label{fig:dynamic_model}
\end{figure}
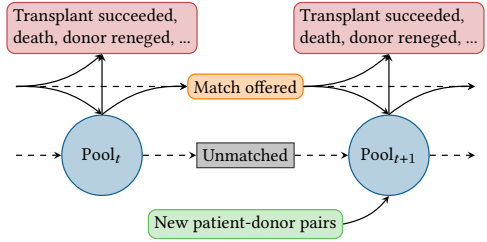

\paragraph{\textbf{Traditional Platforms}}
Typically, at the core of today's traditional kidney exchange platforms (cf.\ Figure~\ref{fig:dynamic_model}) is the pool of those patient-donor pairs that are currently registered and seek for an exchange partner. Usually, a patient-donor pair is associated with a transplant center or hospital which registers it with a central platform by providing them the medical data of both patient and donor. Operators of these central platforms then carry out so-called \emph{match runs} at specific points in time (denoted as $ t $ and $ t + 1 $ in Figure~\ref{fig:dynamic_model}). 

A match run corresponds to solving the \emph{Kidney Exchange Problem} (KEP) among all patient-donor pairs within the pool. The KEP is defined as finding a set of exchanges that maximizes the number of patients that can receive a transplant~\cite{AbrahamClearingAlgorithms2007}. Usually, it is modeled as a graph problem where each patient-donor pair corresponds to one node and an edge is added between two nodes if the donor of the first pair can donate to the patient of the second pair. Exchanges are then computed such that the donor of a patient-donor pair only donates her kidney if the corresponding patient also receives a compatible kidney transplant in return. 

The simplest form of such an exchange is a \emph{crossover exchange} where the donor of an incompatible patient-donor pair $ A $ donates to the patient of another incompatible patient-donor pair $ B $ and vice versa. Larger exchange cycles where the donor of a pair always donates to the patient of the succeeding pair are also possible. However, large cyclic exchanges require a lot of medical resources as all involved transplants have to be carried out simultaneously to prevent a donor from backing out after the corresponding patient already received a transplant. Therefore, most countries only consider exchange cycles of maximum size three (e.g., Netherlands, Spain, UK, and the major exchange platforms in the US~\cite{Biro_EuropeanModellingKE_2019,AshlagiKidneyExchangeOperations2021}) or even only crossover exchanges (e.g., all countries of Scandinavia~\cite{Anderson_PairwiseSweden_2020,Biro_EuropeanModellingKE_2019}).
Besides such cyclic exchanges, also chains initiated by an altruistic donor without a corresponding patient are possible. However, they are still not allowed in many countries~\cite{Biro_EuropeanModellingKE_2019}. 

After the execution of a match run, those pairs that were matched are removed from the pool and informed of the computed exchange partner. If the match results in a transplant, the pairs leave the platform. If the match fails, they reenter the pool. Pairs that were not matched, simply remain in the pool until the next match run is executed. In between two match runs new patient-donor pairs may arrive at the pool and pairs that are already in the pool may leave for various reasons (e.g.,  death or illness of patient or donor).

\paragraph{\textbf{Privacy Concerns}}
The traditional centralized setting exhibits two major shortcomings. First, the patient-donor pairs (or their representative hospitals) have to completely trust the single operator of the centralized platform in computing the exchanges correctly and treating each patient-donor pair equally. Second, an attacker only has to compromise a single entity in order to gain complete control over the exchange computation as well as the sensitive medical data of all patients and donors registered with the platform. 

\paragraph{\textbf{Intuition for the Privacy-Preserving Approach}}
Pursuing a decentralized approach based on \emph{Secure Multi-Party Computation} (SMPC) allows for the distribution of the exchange computation among multiple parties, making it privacy-preserving. In our approach we substitute the central kidney exchange platform by multiple so-called \emph{computing peers} who execute an SMPC protocol among each other to compute the exchanges in a distributed fashion. The patient-donor pairs are then also referred to as \emph{input peers} as they (or their representative hospitals) just send their input (medical data) to the computing peers. They use secret sharing to guarantee that a single computing peer does not gain any knowledge on the actual medical data of any patient or donor. After the protocol execution, the computing peers then send the shares of the computed exchanges to the corresponding input peers. During the protocol execution, the computing peers do not learn anything about the medical data of the patient-donor pairs or the computed exchanges.

Choosing the optimal number of computing peers here means a trade-off between privacy and performance. On one hand, a larger number of computing peers, increases the number of peers that need to be corrupted in order to compromise the privacy of the patient-donor pairs. On the other hand, a smaller number of computing peers decreases the communication overhead induced by the protocol and thus also decreases the protocol runtime. For our runtime measurements (cf. Section~\ref{sub:runtime_measurements}), we use three computing peers which is usually considered a good trade-off~(e.g.,~\cite{Bogdanov_Sharemind_2008,Bogdanov_Estonian_2015}). The computing peers could be governmental institutions, transplant centers, or institutions that cover the patients' interests.

For the actual SMPC protocol that is executed between the computing peers we then focus on pure crossover exchanges. This allows us to reduce the KEP to finding a maximum matching in a general graph where an undirected edge is added between two nodes if the donor of one pair is compatible with the patient of the other and vice versa, i.e., if a crossover exchange between the two pairs is possible. Thus, computing a maximum matching is equivalent to maximizing the number of crossover exchanges. This is a huge advantage since the KEP is NP-complete if the cycle size is restricted to three or larger~\cite{AbrahamClearingAlgorithms2007}. Solving the maximum matching problem on general graphs instead is possible in polynomial time (e.g.,~\cite{EdmondsBlossomAlgorithm1965}). Using an efficient algorithm as basis for a privacy-preserving protocol is important as such protocols in general introduce additional overhead to the original algorithm. Another advantage that comes along with crossover exchanges is that in some countries, e.g., Germany a transplantation is only allowed if patient and donor know each other well which is much easier to guarantee for crossover exchanges than for larger cycles.~\footnote{§8 Transplantationsgesetz (German Transplantation Law)}

\section{Preliminaries}\label{sec:preliminaries}
In this section, we establish the relevant background for our privacy-preserving protocol for kidney exchange including basic terminology from graph theory (Section~\ref{sub:graph_theory}) and a description of the considered setting for secure multi-party computation (Section~\ref{sub:smpc}).

\subsection{Graph Theory}\label{sub:graph_theory}

An \emph{undirected graph} is denoted as $ G = (V,E) $ where $ V $ is a finite set of nodes and $ (u,v) \in E $ indicates an undirected edge between nodes $ u, v \in V $ with $ u \neq v $. Two nodes $ u $ and $ v $ are called \emph{adjacent} if there is an edge $ (u,v) \in E $ and an edge $ (u,v) \in E $ is said to be \emph{incident} to node $ u $ and $ v $. A \emph{matching} is a set of edges $ M $ such that each node $ v \in V $ is incident to at most one edge in $ M $. An edge $ (u, v) \in M $ is called a \emph{matched} edge and an edge $ (u,v) \notin M $ is called an \emph{unmatched} edge. A node $ v \in V $ that is incident to an edge in $ M $ is referred to as a \emph{matched} node and the nodes $ u,v $ of an edge $ (u,v) \in M $ are called \emph{mates}. All nodes $ v \in V $ that are not incident to an edge in $ M $ are called \emph{exposed} nodes. A \emph{path} is a sequence of pairwise different nodes $ p = v_1, ..., v_k $ such that for all $ v_i $ with $ i \in \{1, ..., k-1\} $ it holds that $ (v_i, v_{i+1}) \in E $. A path $ P $ is called \emph{alternating} w.r.t.\ a matching $ M $ if the edges in $ P $ are alternately matched and unmatched edges. We refer to an alternating path $ P $ beginning at an exposed node~$ v_1 $ and ending at an exposed node~$ v_2 $ (with $ v_1 \neq v_2 $) as an \emph{augmenting} path. A \emph{cycle} is a path with $ (v_k, v_1) \in E $. We call a cycle \emph{odd} if it contains an odd number of edges and \emph{even}, otherwise. 
A fundamental property for matching algorithms is Berge's Theorem. 

\begin{definition}{(Berge's Theorem~\cite{BergeTheorem1957})}\label{def:berge}
	\textit{
		A matching $ M $ in a graph~$ G $ is called a maximum matching if and only if $ G $ has no augmenting path with respect to~$ M $.
	}
\end{definition}

Intuitively, Berge's Theorem states that if there is an augmenting path $ P = (v_1, ..., v_k) $ with respect to $ M $ in the graph $ G $, then it is possible to increase the matching by adding all edges $ (v_i, v_{i+1}) \notin M $ with $ i \in \{1, ..., k-1\} $ to $ M $ and removing all edges $ (v_i, v_{i+1}) \in M $ with $ i \in \{1, ..., k-1\} $ from $ M $. Thereby, we obtain a new matching $ M' $ with $ \vert M' \vert = \vert M \vert + 1 $. Thus, $ M $ cannot be a maximum matching.

\subsection{Secure Multi-Party Computation}\label{sub:smpc}

Generally speaking, SMPC allows a set of $ \numCompPeers $ parties to jointly compute a functionality in a distributed fashion such that no party learns anything beyond its private input and output and what can be deduced from both. We adopt the well-established approach (e.g.,~\cite{Bogdanov_Sharemind_2008,Bogdanov_Estonian_2015}) of distinguishing between two sets of parties, so-called input peers and computing peers. While the former provide input to the functionality and receive their corresponding output, the latter execute the actual protocol among each~other. In our protocol (cf.~Section~\ref{sub:protocol_spec}), the input peers correspond to the patient-donor pairs and the computing peers can be governmental institutions, transplant centers, or institutions that cover the patients' interests.

To realize SMPC, we use Shamir's $ (t,n) $ threshold secret sharing scheme~\cite{ShamirSecretSharing1979} which allows the input peers to share a secret value~$ x $ among $ n $ computing peers such that possession of a subset of at most $ t $ shares does not reveal any information on $ x $ itself. Restoring the secret~$ x $ is possible iff at least $ t+1 $ computing peers collaborate. 

We require that all computations are carried out over a finite field $ \mathbb{Z}_p $ for a prime $ p > n $ and we represent a negative value $ -x $ as $ p - x \in \mathbb{Z}_p $. This allows for the correct addition and multiplication of negative values. We denote that a value $ x $ is secret shared among the $ n $ computing peers by $ \enc{x} $. For a vector~$ \enc{V} $ of shared values, we write $ \enc{V}(i) $ to denote the $ i $-th entry of the vector. Analogously, we write $ \enc{M}(i,j) $ for the entry in the $ i $-th row and the $ j $-th column of a matrix $ \enc{M} $ of shared values.

\paragraph{\textbf{Building Blocks}}
Using Shamir's secret sharing scheme enables the computing peers to compute any linear combination of secret shared values locally whereas for multiplication the peers have to interact with each other by means of a multiplication protocol which can be constructed such that it runs in a constant number of rounds~\cite{BenOr_ConstantRoundsMult_1988}. For the sake of readability we use the infix notation $ \enc{x} \leftarrow \enc{y} \cdot \enc{z} $ to denote the execution of a multiplication protocol.
Note that we measure the communication complexity of an SMPC protocol in terms of the number of calls to the multiplication protocol. The round complexity refers to the number of messages transmitted during the SMPC protocol.

In addition to multiplication, we require several other primitives as building blocks for our main protocol (cf.~Section~\ref{sub:protocol_spec}). Secure comparison of two shared values $ \enc{x} $, $ \enc{y} $ is denoted by $ \enc{x} \sharedLessThan \enc{y} $ and can be implemented such that it has linear communication and constant round complexity~\cite{Catrina_PrimitivesSMPC_2010}. Similarly, protocols for equality and inequality test can be constructed. We also require a protocol for conditional selection, i.e., given a secret shared bit $ \enc{b} $, choose $ \enc{x} $ if $ b $ is equal to $ 1 $ and $ \enc{y} $, otherwise. Such a protocol can be realized by computing $ \enc{b} \cdot (\enc{x} - \enc{y}) + \enc{y} $ and we denote it by $ \enc{b} \ ? \ \enc{x} \ : \ \enc{y} $. Finally, we sometimes require to access or update an entry $ \enc{V}(\enc{i}) $ of a vector where the index $ \enc{i} $ is a secret value. Such a secret index is first translated into a vector $ \enc{I} $ containing the value $ 1 $ at position $ i $ and zeros at all other positions. This can be done in linear communication and constant round complexity~\cite{Launchbury_Multiplex_2012}. Based on the index vector $ \enc{I} $, entry $ \enc{V}(\enc{i}) $ can then be accessed by computing the inner product of $ \enc{V} $ and $ \enc{I} $ and updated by computing $ \enc{V}(j) \leftarrow \enc{I}(j) \ ? \ \enc{x} \ : \ \enc{V}(j) $ for $ j \in \{1, \vert V \vert \} $. Both operations require $ \vert \enc{V} \vert $ multiplications which can be executed in parallel. Thus, accessing or updating an entry~$ \enc{V}(\enc{i}) $ overall requires linear communication and constant round complexity.\footnote{This can also be realized with sub-linear complexity using ORAM techniques (e.g.,~\cite{Keller_ORAM_2014}). However, in practice ORAM only provides for a performance improvement for large vector sizes ($ \geq {\sim}1000 $)~\cite{Keller_ORAM_2014} and the vectors in our protocols are much smaller.}

\paragraph{\textbf{Security}}

For our protocol (cf.~Section~\ref{sub:protocol_spec}), we consider security in the \textit{semi-honest model} where the corrupted computing peers strictly follow the protocol specification but try to learn as much as possible on the honest computing peers' input. Security in this sense can be considered sufficient for kidney exchange as it prevents an adversary from learning the medical data of the patient-donor pairs and thus from influencing the computed exchanges in any meaningful way. Note that for the input peers it is guaranteed that they do not learn anything beyond their private input and output as they do not participate in the actual protocol~execution. Furthermore, we assume an honest majority of computing peers as well as encrypted and authenticated channels between the computing peers and between computing peers and input peers.

To prove the security of our main protocol (cf.~Section~\ref{sub:protocol_spec}), we use the standard simulation-based security paradigm which intuitively states that the parties do not learn anything beyond their private input and output. 
For a detailed definition of the security setting and the adversary models, we refer the reader to~\cite{GoldreichFoundationsTwo2004}.

\section{Related Work}\label{sec:related_work}

In this section, we review pertinent related work: state-of-the-art algorithms for maximum matching on general graphs (Section~\ref{sub:matching_algs}); existing privacy-preserving protocols for matching on bipartite graphs (Section~\ref{sub:matching_bipartite}); and the only other privacy-preserving protocol known to date in the context of kidney exchange (Section~\ref{sub:privacyPreservingKE}).

\subsection{Matching Algorithms}\label{sub:matching_algs}

The first efficient algorithm to solve the maximum matching problem on general graphs was devised by Edmonds~\cite{EdmondsBlossomAlgorithm1965} and is based on Berge's Theorem (Definition~\ref{def:berge}). It starts with an initial matching $ M $ which can also be the empty matching. Then, it tries to find an augmenting path $ P $ with respect to $ M $ and revert the path such that a new matching $ M' $ of cardinality $ \vert M \vert + 1 $ is obtained. This is repeated until no further augmenting path can be found. Due to Berge's Theorem the thus obtained matching is a maximum matching.

The major challenge when designing a matching algorithm in this way is the handling of so-called \emph{blossoms}. A blossom is a cycle of odd length in the graph that contains alternating matched and unmatched edges. The problem with blossoms is that it is not trivial how to choose the matching edges of the blossom such that a possibly existing augmenting path is indeed found. Edmonds proposes to shrink a blossom into a single supernode and then continue the algorithm on the shrunken graph. After finding an augmenting path in this adapted graph, the supernode is expanded again and the matching edges of the blossom are chosen accordingly. While this concept is quite simple, the shrinking and expanding of blossoms are computationally rather complex operations leading to a complexity of $ \bigo{\vert V \vert ^4} $ for Edmonds' algorithm. 

The approach by Edmonds has been refined many times by introducing different techniques for handling blossoms, e.g., labeling techniques that allow to avoid the explicit shrinking and expanding of blossoms~(e.g.,~\cite{GabowMatching1976,PapeMatching1980}). The most efficient solutions for the maximum matching problem run in $ \bigo{\sqrt{\vert V \vert} \cdot \vert E \vert} $ time~\cite{Blum_Matching_1990, Gabow_FastMatching_1991,MicaliMatching1980}.

However, for an algorithm to be suitable as the basis for a privacy-preserving protocol it is also important that it has an easy structure and does not rely on complex and dynamically growing datastructures as we have to make sure that the datastructures do not leak any information on the underlying data. 
Therefore, we use the matching algorithm designed by Pape and Conradt~\cite{PapeMatching1980} which uses a very simple labeling technique to avoid the shrinking of blossoms as basis for our privacy-preserving protocol. 
Their approach is based on Edmonds' algorithm and achieves a complexity of $ \bigo{\vert V \vert^3} $.

\subsection{Privacy-Preserving Bipartite Matching}\label{sub:matching_bipartite}
While to the best of our knowledge we are the first to propose a privacy-preserving protocol for maximum matching on general graphs, there are privacy-preserving solutions for maximum matching on bipartite graphs (e.g., \cite{Anandan_Bipartite_2017,Blanton_Bipartite_2015,WuellerHungarianBartering2017}) with complexities ranging from $ O(|V|^3) $ \cite{Blanton_Bipartite_2015} to $ O(|V|^6) $ \cite{WuellerHungarianBartering2017}. Furthermore, the maximum matching problem on bipartite graphs can be translated into the maximum flow problem. Aly et al.~\cite{Aly_Bipartite_2013} present two protocols for the maximum flow problem based on the Edmonds-Karp algorithm and the Push-Relabel algorithm with complexities $ O(|V|^4) $ and $ O(|V|^5) $, respectively. Blanton et al.~\cite{Blanton_Bipartite_2013} compute a maximum flow based on the Ford-Fulkerson algorithm in $ O(|V|^3 |E| log(|V|)) $.

Another related problem is private stable matching where two groups of individuals state their preferences over each other and are matched such that there are no pairs from the two groups who would prefer each other over their computed match. In private stable matching the individuals’ preferences and the outcome of the computation are kept private. The first private stable matching algorithm goes back to Golle~\cite{Golle_Bipartite_2006} and has complexity $ O(|V|^5) $. The~currently best solutions achieve complexity $ O(|V|^2 log^3(|V|)) $~\cite{Doerner_Bipartite_2016,Riazi_Bipartite_2017}.

However, it is not possible to adapt these solutions to crossover kidney exchange as a bipartite graph requires the division of the patient-donor pairs into two sets. The only meaningful way to do so, is to consider donors and patients as two separate sets which makes it impossible to guarantee that a donor only donates her kidney if the corresponding patient also receives a kidney transplant.

\subsection{Privacy-Preserving Kidney Exchange}\label{sub:privacyPreservingKE}

To the best of our knowledge, the only existing privacy-preserving approach for kidney exchange is the SMPC protocol for solving the kidney exchange problem by Breuer et al.~\cite{Breuer_KEprotocol_2020}. The authors rely on SMPC based on additive homomorphic encryption and use a brute force approach to compute a set of exchange cycles that maximizes the number of patients that can receive a kidney transplant. 

While their approach solves the KEP for an arbitrary maximum cycle size, the runtime of their protocol increases very fast for increasing numbers of patient-donor pairs as the KEP is NP-complete for a cycle size larger than 2. Therefore, we restrict our approach to crossover exchange enabling the computation of exchanges between a much larger number of patient-donor pairs.

Another drawback of their approach is that it requires each patient-donor pair to participate in the whole protocol execution which does not reflect a realistic scenario for the use case of kidney exchange since the patient-donor pairs lack the required expertise and the medical data lies with the 
transplant centers or hospitals anyway. Our approach instead only requires the pairs (or hospitals) to send the medical data to three computing peers using secret sharing and then wait for their output. 
\section{Protocol for Crossover Kidney Exchange}\label{sec:protocol}

Before providing the detailed specification of our privacy-preserving protocol for crossover kidney exchange (Section~\ref{sub:protocol_spec}), we introduce the matching algorithm by Pape and Conradt~\cite{PapeMatching1980} on which our protocol is based (Section~\ref{sub:pape_conradt}). In Section~\ref{sub:runtime_measurements}, we present a performance analysis of the implementation of our protocol in MP-SPDZ~\cite{KellerMPSPDZ2020}.

\subsection{Algorithm by Pape and Conradt}\label{sub:pape_conradt}

The main idea of the matching algorithm by Pape and Conradt~\cite{PapeMatching1980} is to develop a blossom (an odd cycle) in two alternating paths instead of shrinking it which is a computationally expensive operation. Thereby, a node that is part of a blossom will be once at even distance along the path from the root and once at odd distance along a second path from the root. Thus, if nodes of the blossom are part of an augmenting path, this path is guaranteed to be found.

To develop a blossom in two alternating paths, an \emph{alternating tree} is grown from each exposed node until either an augmenting path has been found or the tree can be grown no more. The root of an alternating tree is the exposed node~$ r $ which forms level 0 of the tree. All nodes adjacent to $ r $ are added on level 1 and all matched edges incident to the nodes on level 1 (i.e., their mates) form the nodes on level~2. Level~3 then again contains all nodes adjacent to the nodes on level~2 (which are not already part of the tree) and so on. More formally, an alternating tree with respect to a matching~$ M $ is rooted at an exposed node and all paths emanating from the root are alternating paths. The nodes at even levels of the tree are called \emph{inner nodes} and those at odd levels are called \emph{outer nodes}. 

In the algorithm, three datastructures are used to represent an alternating tree. The queue $ \queue $ contains all outer nodes that are currently in the tree but from which the tree has not been explored further. The binary array $ \nontree $ of size $ \vert \nodes \vert $ keeps track of the nodes that are in the tree at even levels (i.e., the root node and all inner nodes). The entry $ \nontree(v) $ is set to $ 0 $ if $ v $ is the root node or an inner node and to $ 1 $ if it is an outer node. The array $ \grandfather $ of size~$ \vert \nodes \vert $ is used to trace back the path up to the root if an augmenting path has been found. For an outer node $ w $, $ \grandfather(w) = u $ indicates that there is a path $ (u,v,w) $ in the alternating tree from $ u $ to $ w $ such that $ (v,w) $ is a matched edge. 

\begin{algorithm}[!t]
	\caption{Matching Algorithm by Pape and Conradt~\cite{SysloMatchingBook1983}}
	\label{alg:pape}
	\algrenewcommand\algorithmicindent{1em}
\small
\begin{algorithmic}[1]
	\State Start with an initial matching
	\For{$ r \in \nodes $}
		\If{$ \mate(r) = 0 \wedge \expo \geq 2 $}
			\For{$ v \in \nodes $}
				\State $ \nontree(v) \gets 1 $
			\EndFor
			\State $ \nontree(r) \gets 0 $
			\State $ \queue.\textit{insert(r)} $
			\State $ \found \gets 0 $
			\While{$ \queue \neq \emptyset \wedge \found = 0 $}
				\State $ x \gets \queue.\textit{removeHead()} $
				\For{$ y $ adjacent to $ x $}
					\If{$ \nontree(y) = 1 $}
						\If{$ \mate(y) = 0 $}
							\State Update matching
							\State $ \expo \gets \expo - 2 $
							\State $ \found \gets 1 $
						\ElsIf{$ \mate(y) \neq x $}	
							\If{$ y $ is not an ancestor of $ x $}
								\State $ z \gets \mate(y) $
								\State $ \nontree(y) \gets 0 $
								\State $ \grandfather(z) \gets x $
								\State $ \queue.\textit{insert(z)} $
							\EndIf
						\EndIf
					\EndIf
				\EndFor	
			\EndWhile
		\EndIf
	\EndFor
\end{algorithmic}
\end{algorithm}

Algorithm~\ref{alg:pape} contains pseudocode for the matching algorithm by Pape and Conradt~\cite{PapeMatching1980}. The algorithm starts with an initial matching which can be any matching $ M $ in $ G $ including the empty matching. The matching is given by the array $ \mate $ of size $ \vert \nodes \vert $ which encodes the mate of each node~$ v \in \nodes $. The number of exposed nodes is given by $ \expo $. The algorithm iterates over all potential root nodes $ r \in V $ and starts growing an alternating tree from $ r $ if $ r $ is an exposed node. The tree is grown based on a breadth-first search. Note that if the number of exposed nodes is less than 2, the matching is already a maximum matching and the algorithm can be aborted. 

If $ M $ is not yet a maximum matching, the array $ \nontree $ is initialized such that only the entry for the root node $ r $ equals~$ 0 $ indicating that the tree currently only contains $ r $. Also, $ r $ is added to the queue~$ \queue $ of outer nodes to be explored from and the boolean $ \found $ is set to $ 0 $ stating that an augmenting path has not yet been found. 

Then, the first node $ x $ is removed from $ \queue $ (which is the root node~$ r $ in the first iteration) and a tree is grown from $ x $. The algorithm parses over each node $ y $ that is adjacent to $ x $. If $ \nontree(y) = 0 $, the edge $ (x,y) $ is ignored as $ y $ is an inner node, i.e., it is already part of the tree. If $ \nontree(y) = 1 $, the node is either matched or exposed. If $ y $ is exposed (i.e., $ \mate(y) = 0 $), an augmenting path from root node~$ r $ to $ y $ has been found. In that case, the matching $ M $ is updated, the number of exposed nodes is decreased by 2, the current tree is abandoned, and an alternating tree is grown from the next exposed node. If $ y $ is not exposed and the mate of $ y $ is not $ x $ itself, it has to be checked whether $ y $ is an ancestor of $ x $ in the tree. If $ y $ is an ancestor of $ x $, they are part of a blossom (an odd cycle) which could lead to a false augmenting path. In that case the algorithm just continues with the next value for $ y $.  If $ y $ is not an ancestor of $ x $, the inner node~$ y $ and the outer node~$ z = \mate(y) $ are added to the tree, i.e., $ z $ is added to $ \queue $, $ \grandfather(z) $ is set to $ x $, and $ \nontree(y) $ is set to $ 0 $. 

A maximum matching has been found if there are less than two exposed nodes or if every exposed node has been tried as the root of an alternating tree. 
The complete algorithm has complexity~$ \bigo{\vert V \vert^3} $. For a detailed complexity analysis, we refer the reader to~\cite{SysloMatchingBook1983}.

\subsection{SMPC Protocol Specification}\label{sub:protocol_spec}

Our privacy-preserving protocol for kidney exchange is based on the matching algorithm by Pape and Conradt~\cite{PapeMatching1980} (in the following referred to as conventional PC algorithm) introduced in Section~\ref{sub:pape_conradt}. Specifically, we first have to construct the graph on which we compute the matching such that each node corresponds to one patient-donor pair and an undirected edge is added between two nodes if the corresponding pairs are medically compatible, i.e., the donor of one pair can donate to the patient of the other and vice versa. Since each edge then corresponds to one crossover exchange, computing a maximum matching on the constructed graph then is equivalent to maximizing the number of crossover exchanges. 

The detailed specification of protocol $ \crossoverKeProtocol $ for privacy-preserving crossover kidney exchange is given in Protocol~\ref{prot:matching_short}. Note that prior to the protocol execution, all input peers (patient-donor pairs) send their input (medical data) to the computing peers using secret sharing. Thus, each computing peer holds a share of the medical data of each patient-donor pair $ \partysymbol{v} $ with $ v \in \nodes $ where $ \vert \nodes \vert $ corresponds to the number of patient-donor pairs for which the protocol is executed. 

\paragraph{\textbf{Graph Initialization Phase}}
At the beginning of the protocol, a shared adjacency matrix $ \enc{\compGraphMatrix} $ is computed such that an entry $ \enc{\compGraphMatrix}(u,v) $ encodes whether patient-donor pairs $ \partysymbol{u} $ and $ \partysymbol{v} $ ($ u,v \in \nodes $) are compatible. To this end, we use the building block~$ \compCheck $ introduced by Breuer et al.~\cite{Breuer_KEprotocol_2020}. The input to the compatibility check comprises secret binary indicator vectors for the donor bloodtype~$ \enc{\donorbloodvecNotSpec} $, the patient bloodtype~$ \enc{\patientbloodvecNotSpec} $, the donors' antigens~$ \enc{\antigenvecNotSpec} $, and the patients' antibodies~$ \enc{\antibodyvecNotSpec} $ for both patient-donor pairs. The donor of pair~$ \partysymbol{u} $ is then considered compatible with the patient of pair~$ \partysymbol{v} $ if $ \donorbloodvec{u}(i) = \patientbloodvec{v}(i) = 1 $ for at least one $ i \in \{1, ..., \vert \donorbloodvecNotSpec \vert \} $ and if for all $ i \in \{1, ..., \vert \antibodyvecNotSpec \vert \} $ with $ \antigenvec{u}(i) = 1 $ it holds that $ \antibodyvec{v}(i) \neq 1 $, i.e., if the patient has no antibody against the donor's antigens.\footnote{Additional criteria for compatibility can be checked in the same way if required.} As compatibility has to be computed between the donor of pair $ \partysymbol{u} $ and the patient of pair~$ \partysymbol{v} $ and vice versa, this requires $ 2 \cdot \vert \donorbloodvecNotSpec \vert \cdot \vert \antigenvecNotSpec \vert $ comparisons which can be executed in parallel. Note that even if the protocol indicates that two pairs are compatible, the final choice of whether a transplant is carried out still lies with medical experts.
Before starting the matching computation, the adjacency matrix is shuffled at random to ensure that two pairs with the same input have the same probability of being matched independent of their index. It is sufficient that $ \frac{\numCompPeers}{2} + 1 $ of the $ \numCompPeers $ computing peers each input a secret permutation matrix. All rows and columns of the matrix are then shuffled once with each of these permutation matrices.\footnote{The graph initialization phase is the only part of protocol~$ \crossoverKeProtocolFootnote $ specific to kidney exchange. To consider a different use case, one just has to change the~computation of the adjacency matrix. Thus, our protocol is of independent interest for any use case requiring the privacy-preserving computation of a matching on general~graphs.}

\paragraph{\textbf{Path Finding Phase}}
After the computation of the adjacency matrix, we initialize an empty matching by setting the mate of each patient-donor pair~$ \partysymbol{u} $ ($ \forall u \in \nodes $) to $ \enc{0} $. In the privacy-preserving setting the initial matching does not influence the protocol performance as the worst case number of iterations have to be executed for all loops in order to hide the structure of the alternating trees.

\begin{protocol}[t]
	\caption{$ \crossoverKeProtocol $}
	\label{prot:matching_short}
	\algrenewcommand\algorithmicindent{1em}
\small
\begin{algorithmic}[1]

	\For{$ u,v \in \nodes $}
		\State $ \enc{A}(u,v) \leftarrow \compCheck(\enc{\donorbloodvec{u}}, \enc{\antigenvec{u}}, \enc{\patientbloodvec{u}}, \enc{\antibodyvec{u}},$\newline
		\hspace*{12.2em} $ \enc{\donorbloodvec{v}}, \enc{\antigenvec{v}}, \enc{\patientbloodvec{v}}, \enc{\antibodyvec{v}}) $
	\EndFor
	\State $ \enc{\compGraphMatrix} \leftarrow \shuffleGate(\enc{\compGraphMatrix}) $
	\For{$ u \in \nodes $}
		\State $ \enc{\mate}(u) \leftarrow \enc{0} $
	\EndFor

	\For{$ r \in \nodes $}
		\State $ \enc{\rootNode} \leftarrow (\enc{\mate}(r) \sharedEqualityCheck \enc{0}) \ ? \ \enc{r} \ : \ \enc{0}) $
		\For{$ v \in \nodes $}
			\State $ \enc{\nontree}(v) \leftarrow \enc{1} $
		\EndFor
		\vspace*{-0.4em}
		\State $ \enc{\nontree}(r) \leftarrow \enc{\rootNode} \sharedEqualityCheck \enc{0} $ 
		\State $ \enc{\queue} \leftarrow \queueInsert{\enc{\rootNode}} $ 
		\State $ \enc{\found} \leftarrow \enc{0} $
		\For{$ \vert \nodes \vert $ times}
			\State $ \enc{x} \leftarrow \queueRemove $
			\State $ \enc{x} \leftarrow \enc{\found} \ ? \ \enc{0} \ : \ \enc{x} $
			\For{$ y \in \nodes $}
				\State $ \enc{\foundPath} \leftarrow \enc{\nontree}(y) \sharedMult \enc{A}([x], y) \sharedMult $ \newline 
				\hspace*{8.6em} $ (\enc{\mate}(y) \sharedEqualityCheck \enc{0}) \sharedMult (\enc{\mate}(r) \sharedEqualityCheck \enc{0}) $

				\State $ \enc{\found} \leftarrow \enc{\foundPath} \sharedAdd \enc{\found} \sharedSubtract \enc{\foundPath} \sharedMult \enc{\found} $
				
				\State $ \enc{\mate} \leftarrow \enlargeMatchingGate(\enc{\foundPath}, \enc{\mate}, \enc{\textit{grandf.}}, \enc{x}, y) $

				\State $ \enc{\zAncX} \leftarrow \checkOddCycleGate(\enc{\grandfather}, \enc{x}, y) $ 
				\State $ \enc{\textit{c}_2} \leftarrow \enc{\nontree}(y) \sharedMult (\enc{x} \sharedUnequalityCheck \enc{\mate}(y)) \sharedMult $ \newline
				\hspace*{6.1em} $ (\enc{\found} \sharedEqualityCheck \enc{0}) \sharedMult \enc{A}([x], y) \sharedMult (\enc{\zAncX} \sharedEqualityCheck \enc{0}) $
				\State $ \enc{z} \leftarrow \enc{\textit{c}_2}  \ ? \ \enc{\mate}(y) \ : \ \enc{0}) $
				\State $ \enc{\nontree}(y) \leftarrow (\enc{z} \sharedUnequalityCheck \enc{0}) \ ? \ \enc{0} \ : \ \enc{\nontree}(y) $
				\State $ \enc{\grandfather}([z]) \leftarrow \enc{x} $ 
				\State $ \enc{\queue} \leftarrow \queueInsert{\enc{z}} $
			\EndFor
		\EndFor
	\EndFor 
	\State $ \enc{\mate} \leftarrow \revShuffleGate(\enc{\mate}) $
	\For{$ v \in \nodes $}
		\State Send share of $ \enc{\mate}(v) $ to patient-donor pair $ \partysymbol{v} $
	\EndFor
\end{algorithmic}
\end{protocol}

The computation of a maximum matching then starts with the for-loop over all potential root nodes $ r $. In contrast to the conventional PC algorithm, we cannot abort the tree construction if the node $ r $ is not exposed as we have to keep all information about the structure of the underlying graph private. Thus, we set $ \enc{\rootNode} $ to $ \enc{0} $ if $ r $ is not exposed, i.e., all relevant conditions in the remainder of the loop will also be equal to $ \enc{0} $ 
and thus the matching will not be altered for $ \enc{\rootNode} = \enc{0} $. Note that the entire protocol is designed such that if a node $ v $ has value $ 0 $, no further modifications to the current state of the tree are made (i.e., we use $ 0 $ as a \emph{dummy node}). 

Then, we initialize the array $ \enc{\nontree} $, the queue $ \enc{\queue} $, and the boolean $ \enc{\found} $ as in the conventional PC algorithm. If $ \enc{\rootNode} = \enc{0} $, $ \enc{\nontree}(r) $ is set to $ \enc{1} $ and $ \enc{0} $ is inserted into the queue~$ \enc{\queue} $.

In the conventional PC algorithm, one would now iterate over the entries in the queue $ \enc{\queue} $ until it is empty or an augmenting path is found. However, to avoid any information leakage w.r.t.\ the structure of the alternating tree, we have to execute the loop for a fixed number of iterations. In particular, we have to execute the loop for the worst case number of iterations which is $ \vert \nodes \vert $ as each node can be added at most once to the queue~$ \enc{\queue} $. 

The loop then starts with the first node $ \enc{x} $ from the queue~$ \enc{\queue} $. In the privacy-preserving setting, the queue has to be implemented such that it returns $ \enc{0} $ if it is empty. Thus, if the queue is empty, the protocol continues computing on the dummy node $ 0 $.

In the next step, we check whether we have already found an augmenting path for this tree (i.e., $ \enc{\found} = \enc{1} $). If that is the case, we set $ \enc{x} $ to $ \enc{0} $. Otherwise, $ \enc{x} $ remains unchanged.

In the conventional PC algorithm, one would now only iterate over those nodes that are adjacent to $ \enc{x} $. However, as we have to keep the nodes that are indeed adjacent to $ \enc{x} $ hidden, we have to iterate over all nodes $ y \in \nodes $. First, we check if there is an augmenting path rooted at $ r $ and ending in $ y $. This is the case if $ y $ is still not part of the tree (i.e., $ \enc{\nontree} = \enc{1} $), $ y $ is adjacent to $ \enc{x} $ (i.e., $ \enc{\compGraphMatrix}(\enc{x}, y) = \enc{1} $), and $ y $ and $ r $ are both exposed nodes (i.e., $ \enc{\mate}(y) = \enc{\mate}(r) = \enc{0} $). If all these conditions hold, their product equals $ \enc{1} $ and thus $ \enc{\foundPath} $ is set to $ \enc{1} $. Afterwards, we can update the secret variable $ \enc{\found} $ which is used in the next round to check whether we have already found an augmenting path for the tree rooted at $ r $. Furthermore, we call the protocol $ \enlargeMatchingGate $ to increase the matching along the found augmenting path. 

\paragraph{\textbf{Matching Update Phase}}
The specification of protocol $ \enlargeMatchingGate $ is given in Protocol~\ref{prot:enlarge_matching}. The general idea of the protocol is to trace back the path from the exposed node $ y $ to the root node $ r $ and change all matched edges to unmatched edges and vice versa. Thereby, the cardinality of the matching is increased by exactly $ 1 $. Of course, this should only be done if we indeed found an augmenting path, i.e., $ \enc{\foundPath} = \enc{1} $. To this end, at the beginning of protocol~$ \enlargeMatchingGate $ $ \enc{\mate}(y) $ is set to $ \enc{x} $ if $ \enc{\foundPath} = \enc{1} $. Afterwards, we iterate over the edges of the path and switch matched with unmatched edges. We use an auxiliary variable $ \Next $ which stores the next node on the path. If $ \enc{\foundPath} = \enc{1} $, we set $ \enc{\Next} $ to the mate of $ \enc{x} $ and thus to the next node on the path from $ y $ to $ r $. Furthermore, we set the new mate of $ \enc{x} $ to $ \enc{y} $ and $ \enc{x} $ itself to the grandfather of $ \enc{x} $. Then, we set the new mate of $ \enc{\Next} $ to the next value for $ \enc{x} $, i.e., to the grandfather of $ \enc{x} $. Furthermore, we update $ \enc{y} $ such that it now contains the next node to consider on the path (i.e., $ \enc{\Next} $). Thereby, $ \enc{y} $ and $ \enc{x} $ become the two next nodes on the path to the root. We have to execute the whole loop $ \vert \nodes \vert $ times as this is the maximum number of nodes of the augmenting path. Note that $ \enlargeMatchingGate $ does not alter the matching if $ \enc{x} = \enc{0} $ or $ \enc{\foundPath} = \enc{0} $.

\begin{protocol}[!t]
	\caption{$ \enlargeMatchingGate $}
	\label{prot:enlarge_matching}
	\algrenewcommand\algorithmicindent{1em}
\small
\begin{algorithmic}[1]
	\State $ \enc{\mate}(y) \leftarrow \enc{\foundPath} \ ? \ \enc{x} \ : \ \enc{\mate}(y) $
	\For{$ \vert \nodes \vert $ times}
		\State $ \enc{\Next} \leftarrow \enc{\foundPath} \sharedMult \enc{\mate}(\enc{x}) $
		\State $ \enc{\mate}(\enc{x}) \leftarrow \enc{\foundPath} \ ? \ \enc{y} \ : \ \enc{\mate}(\enc{x}) $
		\State $ \enc{x} \leftarrow \enc{\foundPath} \sharedMult \enc{\grandfather}(\enc{x}) $
		\State $ \enc{\mate}(\enc{\Next}) \leftarrow \enc{\foundPath} \sharedMult \enc{x} $
		\State $ \enc{y} \leftarrow \enc{\foundPath} \sharedMult \enc{\Next} $
	\EndFor
\end{algorithmic}

\end{protocol}

\paragraph{\textbf{Tree Extension Phase}}
In the last part of the tree growing in Protocol~\ref{prot:matching_short}, we check if the nodes $ y $ and $ \enc{z} $ ($ \enc{z} $ is the mate of $ y $) have to be added to the tree. First, we check if $ y $ is an ancestor of $ \enc{x} $ in the tree rooted at $ r $ using the protocol~$ \checkOddCycleGate $. If this is the case, the two nodes are part of an odd cycle and should not be added to the tree as they could lead to a false augmenting path.

The specification of $ \checkOddCycleGate $ is given in Protocol~\ref{prot:odd_cycle}. In each iteration, we check whether the current node $ \enc{x} $ is equal to $ y $. If this is the case, we know that $ \enc{x} $ is an ancestor of $ y $. Otherwise, we update the value of $ \enc{x} $ to its grandfather. If we reach the root, $ \enc{x} $ is set to $ \enc{0} $ as $ \enc{\grandfather}(r) = \enc{0} $. Thus, at the end of the protocol $ \enc{\zAncX} = \enc{1} $ if and only if $ y $ is an ancestor of $ \enc{x} $.

Returning to Protocol~\ref{prot:matching_short}, we add $ \enc{z} $ and $ \enc{y} $ to to the tree if all of the following conditions hold: $ y $ is not part of the tree already (i.e., $ \enc{\nontree}(y) = 1 $), $ \enc{\mate}(y) $ is not equal to $ \enc{x} $, no path has been found in the current tree so far (i.e., $ \enc{\found} = \enc{0} $), $ y $ is adjacent to $ \enc{x} $ (i.e., $ \enc{\compGraphMatrix}(\enc{x}, y) = \enc{1} $), and $ y $ is not an ancestor of $ \enc{x} $. Otherwise, we set $ \enc{z} $ to $ \enc{0} $. Afterwards, we update $ \enc{\nontree} $, $ \enc{\grandfather} $, and $ \enc{\queue} $ accordingly, i.e., if $ \enc{z} = \enc{\mate}(y) $, we set $ \enc{\nontree}(y) $ to $ \enc{0} $, $ \enc{\grandfather}(z) $ to $ \enc{x} $, and add $ \enc{z} $ to $ \enc{\queue} $.  

\paragraph{\textbf{Output Phase}}
At the end of the protocol, the entries in $ \enc{\mate} $ are unshuffled using the inverses of the permutation matrices that were used to shuffle the adjacency matrix during the graph initialization phase. 
Afterwards, each patient-donor pair $ \partysymbol{v} $ with $ v \in \nodes $ obtains the shares of $ \enc{\mate}(v) $ indicating its exchange partner.

\paragraph{\textbf{Correctness}} The correctness of Protocol~\ref{prot:matching_short} follows from the correctness of the conventional PC algorithm and the fact that computations on the dummy node $ 0 $ do not alter the computed~matching.

\paragraph{\textbf{Security}} Since throughout Protocol~\ref{prot:matching_short} all computations are executed on shared values and all building blocks are called on shared inputs, the simulator can just call the simulator of the corresponding building block on random input for each building block that is executed during protocol~$ \crossoverKeProtocol $. Security then follows from the properties of Shamir's secret sharing scheme, the security of the primitives introduced in Section~\ref{sub:smpc}, and the data-obliviousness of Protocol~\ref{prot:matching_short}. The latter refers to the property that no information on the structure of the trees is leaked as always all possible branches of a tree are considered (using dummy iterations if an augmenting path has already been found or the queue is empty).

\begin{protocol}[t]
	\caption{$ \checkOddCycleGate $}
	\label{prot:odd_cycle}
	\algrenewcommand\algorithmicindent{1em}
\small
\begin{algorithmic}[1]
	\State $ \enc{\zAncX} \leftarrow \enc{0} $
	\For{$ \left\lceil \vert \nodes \vert / 2 \right\rceil $ times}
		\State $ \enc{\zAncX} \leftarrow (\enc{x} \sharedEqualityCheck \enc{y}) \ ? \ \enc{1} \ : \ \enc{\zAncX} $
		\State $ \enc{x} \leftarrow \enc{\grandfather}(\enc{x}) $
	\EndFor
\end{algorithmic}
\end{protocol}

\paragraph{\textbf{Complexity}} The computation of the adjacency matrix can be done in $ \bigo{\vert \nodes \vert^3} $ communication and $ \bigo{\vert \nodes^2 \vert} $ round complexity as protocol~$ \compCheck $ exhibits linear communication and constant round complexity. The most time consuming part of protocol~$ \crossoverKeProtocol $, however, is the construction of the alternating trees for all potential root nodes. The corresponding loop is executed $ \vert \nodes \vert $ times as are the inner loops in lines~12 and~15 of Protocol~\ref{prot:matching_short}. Note that it is not possible to abort these loops early as this would leak information on the underlying graph. Protocols $ \enlargeMatchingGate $ and $ \checkOddCycleGate $ both have communication complexity $ \bigo{\vert \nodes^2 \vert} $ and round complexity $ \bigo{\vert \nodes \vert} $ and are called once in the innermost loop. Thus, the construction of the alternating trees has communication complexity $ \bigo{\vert \nodes^5 \vert} $ and round complexity $ \bigo{\vert \nodes \vert^4} $ which are also the complexities for the complete protocol~$ \crossoverKeProtocol $.

\subsection{Performance Analysis}\label{sub:runtime_measurements}

We have implemented our novel protocol~$ \crossoverKeProtocol $ in the SMPC benchmarking framework MP-SPDZ~\cite{KellerMPSPDZ2020} and evaluated its performance. The source code is published in~\cite{Breuer_Code_2022}.

In our evaluation, we use LXC containers on a single server with an AMD EPYC 7702P 64-core processor for the computing peers and the input peers. Each container runs Ubuntu 20.04 LTS and is equipped with one core and 4GB RAM. Each computing peer corresponds to a single container and a fourth container manages sending input and receiving output for all input peers (patient-donor pairs). For all measurements where the runtime of a single protocol execution is less than one hour, we execute ten repetitions and average the results. For all others, we only execute the protocol once as deviations between the different runs are comparably low for such long runtimes. As performance measures we consider the runtime and the induced network traffic which corresponds to the accumulated network traffic of all three computing peers. We start a measurement when the computing peers have received all inputs and end it when they have sent the output back to the input~peers. 

Since all containers are hosted on the same server, the \emph{latency~$ L $} in our setup is very low. In order to mimic real-world scenarios, we include the capability to artificially set the latency and the bandwidth to realistic values. As these may differ considerably depending on the setup of the three computing peers in practice, we have to determine values that resemble pertinent use cases. With respect to the bandwidth, it seems reasonable to assume that the computing peers have a high-end Internet connection as they are hosted by large (possibly governmental) institutions. Thus, we assume a bandwidth $ B = 1\textit{Gbps} $ for all our measurements.
\footnote{Results for smaller bandwidths can be found in Appendix~\ref{appendixA}.} 
Concerning the latency, the values may differ considerably depending on the location of the different institutions that host the computing peers. Therefore, we consider three different values for the latency $ L $. For a best-case scenario, where all institutions are hosted in the same network, we consider latency $ \textit{L} = 1\textit{ms} $. As an average case latency, we use $ \textit{L} = 5\textit{ms} $, and as a worst-case setup, we use~$ \textit{L} = 10\textit{ms} $.   

\begin{table}[t]
	\caption{Runtime and network traffic of protocol~$ \crossoverKeProtocol $ for different numbers of input peers and different latencies $ \textit{L} $ for a fixed bandwidth $ \textit{B} = 1 \textit{Gbps} $.\label{tab:runtimes}}
	\centering
\begin{tabular}{c | r  r  r | r}
	\multirow{2}{*}{\centering input peers} & \multicolumn{3}{c|}{runtime}  & \multicolumn{1}{c}{\multirow{2}{*}{traffic}}  \\ \cline{2-4}
			& \multicolumn{1}{c|}{$ L = 1\textnormal{ms} $}	& \multicolumn{1}{c|}{$ L = 5\textnormal{ms} $}	& \multicolumn{1}{c|}{$ L = 10\textnormal{ms} $}	& 			\Tstrut\\ \hline
	5		& 21s			& 97s			& 192s			& 51MB		\Tstrut\\  
	10		& 6m			& 27m			& 53m			& 759MB		\\ 
	15		& 30m			& 2h			& 5h			& 4GB		\\ 
	20		& 96m			& 7h			& 15h			& 13GB		\\
	25		& 4h			& 18h			& 35h			& 33GB		\\
	30		& 8h			& 37h			& 73h			& 70GB		\\
	35		& 15h			& 68h			& 134h			& 148GB		\\
	40		& 25h			& 116h			& -				& 260GB		\\
	45		& 40h			& -				& -				& 423GB		\\
	50		& 61h			& -				& -				& 663GB		\\
	55		& 89h			& -				& -				& 990GB		\\
	60		& 127h			& -				& -				& 1440GB	\\
\end{tabular}

\end{table}

Table~\ref{tab:runtimes} shows runtime and network traffic for increasing numbers of input peers. We stopped a measurement if it took longer than $ 7 $ days.\footnote{We decided to only allow protocol runtimes for up to $ 7 $ days as for longer runtimes the state of the patient-donor pairs inside the pool of a kidney exchange platform may already have changed substantially before the actual protocol execution is finished.} Let us first consider the runtimes for $ \textit{L} = 1\textit{ms} $. We observe that the runtime increases considerably for increasing numbers of input peers. For up to $ 5 $ input peers, the protocol still finishes within one minute for $ \textit{L} = 1\textit{ms} $, whereas the runtime for $ 20 $ input peers amounts to more than $ 1 $ hour and for $ 40 $ input peers the protocol requires more than $ 24 $ hours to finish. Finally, for $ 60 $ input peers the protocol runs for $ 5 $ days and $ 7 $ hours. The network traffic scales similarly: for up to $ 5 $ input peers it is still below $ 50\textit{MB} $, for $ 20 $ input peers it increases to more than $ 10\textit{GB} $, for $ 40 $ input peers $ 260\textit{GB} $, and for $ 60 $ input peers $ 1.4\textit{TB} $ are sent in one protocol run. 

We also observe that the runtime increases for increasing latencies. On average the runtime for $ L = 1\textit{ms} $ is about $ 4.66 $ times larger than for $ L = 5\textit{ms} $ and about $ 9.22 $ times larger than for $ L = 10\textit{ms} $.

\begin{figure}[!t]
	\includegraphics[scale=1]{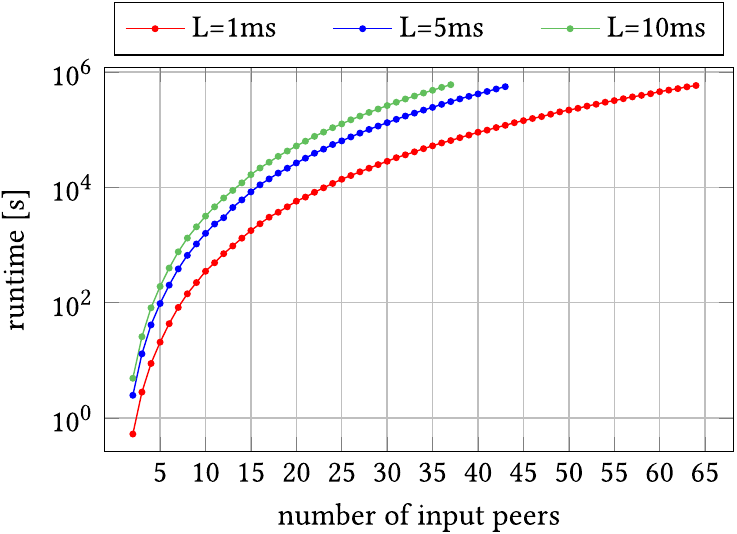}
	\caption{Runtime of protocol~$ \crossoverKeProtocol $ for different latencies $ \textit{L} $ and fixed bandwidth $ \textit{B} = 1 \textit{Gbps} $.}
	\label{fig:runtimes}
\end{figure}

Figure~\ref{fig:runtimes} shows plots of the runtime for the three considered latencies. Note that the scale of the y-axis is logarithmic. We observe that the overall runtime is less than exponential. Furthermore, the plots show that the factor between the runtimes for the different latencies is nearly constant. This is as expected since the latency is a constant offset by which each sent message is delayed.

While the runtimes of Protocol~\ref{prot:matching_short} may seem large, this is not necessarily an issue for our use case as most kidney exchange platforms typically compute a matching only once every couple of days and some even only once every three months~\cite{AshlagiKidneyExchangeOperations2021}. In Section~\ref{sec:simulation}, we analyze to what extent the runtimes of Protocol~\ref{prot:matching_short} allow for its application in large scale kidney exchange platforms.

\subsubsection{Comparison to Related Work}\label{subsub:runtime_comparison}

We compare our novel protocol $ \crossoverKeProtocol $ to the privacy-preserving protocol for kidney exchange by Breuer et al.~\cite{Breuer_KEprotocol_2020} (cf.\ Section~\ref{sub:privacyPreservingKE}). While their protocol allows to solve the KEP for any maximum cycle size, it is also possible to restrict it to pure crossover exchange. To enable a fair comparison between the two approaches, we implemented the protocol from~\cite{Breuer_KEprotocol_2020} in MP-SPDZ considering the same setting as for our protocol~$ \crossoverKeProtocol $ (secret sharing is used instead of homomorphic encryption and the protocol is run by three computing peers whereas the patient-donor pairs correspond to the input peers). We refer to the implementation from~\cite{Breuer_KEprotocol_2020} based on homomorphic encryption as protocol~$ \breuerHomEncProt $ and to our implementation of the protocol from~\cite{Breuer_KEprotocol_2020} based on secret sharing as protocol~$ \breuerSecSharingProt $. We publish the source code for protocol~$ \breuerSecSharingProt $ in~\cite{Breuer_Code_2022}.

Table~\ref{tab:runtimes_compared} shows the results for our protocol $ \crossoverKeProtocol $ and the two protocols $ \breuerSecSharingProt $ and $ \breuerHomEncProt $ (restricted to crossover exchange).\footnote{We only evaluated the protocol~$ \breuerHomEncProtFootnote $ for up to $ 10 $ parties as this already showed that the other two protocols drastically outperform protocol~$ \breuerHomEncProtFootnote $.} We observe that the runtime of protocol~$ \breuerHomEncProt $ is worse than for both other protocols for all numbers of input peers. In particular, by using secret sharing instead of homomorphic encryption we improved the performance of the protocol from~\cite{Breuer_KEprotocol_2020} significantly. When comparing our novel protocol~$ \crossoverKeProtocol $ to the protocol~$ \breuerSecSharingProt $, we observe that for up to $ 12 $ input peers, protocol~$ \breuerSecSharingProt $ outperforms our protocol~$ \crossoverKeProtocol $. However, for larger numbers of input peers, our protocol~$ \crossoverKeProtocol $ is much more efficient. We prescribe this to the brute-force nature of the approach from~\cite{Breuer_KEprotocol_2020}, i.e., an exhaustive search is run over the set of all possible exchange constellations that can exist between the input peers. Since this set increases very fast for increasing numbers of input peers, their approach is only feasible for small numbers of input peers (e.g., for $ 15 $ input peers protocol~$ \breuerSecSharingProt $ already takes $ 17 $ times longer than our protocol~$ \crossoverKeProtocol $).

Our results show that for countries where only crossover exchange is allowed our specialized protocol for crossover exchange is better suited than the existing approach from~\cite{Breuer_KEprotocol_2020}. This was to be expected as the general KEP is an NP-complete problem~\cite{AbrahamClearingAlgorithms2007}. 

\begin{table}[!t]
	\caption{Runtime and network traffic for our protocol $ \crossoverKeProtocol $ (for latency $ \textit{L} = 1\textit{ms} $), our implementation $ \breuerSecSharingProt $ of the protocol from \cite{Breuer_KEprotocol_2020} (for latency $ \textit{L} = 1\textit{ms} $), and the implementation $ \breuerHomEncProt $ from~\cite{Breuer_KEprotocol_2020} where all parties are connected by a LAN with latency $ \textit{L} < 0.5\textit{ms} $.\label{tab:runtimes_compared}}
	\begin{adjustbox}{width=\columnwidth}
	\centering
	\begin{tabular}{c | r r | r r | r r}
		\multirow{1}{*}{input} 	& \multicolumn{2}{c |}{$ \crossoverKeProtocol $}	& \multicolumn{2}{c|}{$ \breuerSecSharingProt $} & \multicolumn{2}{c}{$ \breuerHomEncProt $}	\\ \cline{2-7}
		\multirow{1}{*}{peers}	& \multicolumn{1}{c}{runtime}	& \multicolumn{1}{c|}{traffic}	& \multicolumn{1}{c}{runtime}	& \multicolumn{1}{c|}{traffic} & \multicolumn{1}{c}{runtime}	& \multicolumn{1}{c}{traffic}	\Bstrut\Tstrut\\ \hline
		2 	& 0.5s	& 4MB	& 0.2s	& 0.4MB		& 14s	& 1MB\Tstrut\\
		3 	& 3s	& 10MB	& 0.7s	& 0.8MB		& 23s	& 4MB	\\
		4 	& 9s	& 24MB	& 2s	& 2MB		& 36s	& 11MB	\\
		5 	& 21s	& 51MB	& 3s	& 3MB		& 64s	& 32MB	\\
		6 	& 43s	& 101MB	& 4s	& 6MB		& 136s	& 93MB	\\
		7 	& 83s	& 186MB & 5s	& 12MB		& 6m	& 307MB	\\
		8 	& 2m	& 318MB	& 8s	& 39MB		& 19m	& 1GB	\\
		9 	& 4m	& 492MB	& 13s	& 147MB		& 66m	& 5GB	\\
		10 	& 6m	& 759MB	& 30s	& 585MB		& 4h	& 23GB	\\
		11 	& 8m	& 1GB	& 2m	& 2GB		& - 	& -  	\\
		12 	& 12m	& 1.5GB & 7m	& 10GB		& - 	& - 	\\
		13 	& 16m	& 2GB 	& 26m	& 40GB		& - 	& - 	\\
		14 	& 22m	& 3GB 	& 115m	& 166GB		& - 	& - 	\\
		15 	& 30m	& 4GB 	& 8.5h	& 684GB		& - 	& - 	\\
	\end{tabular}
\end{adjustbox}
\end{table}

\section{Dynamic Kidney Exchange}\label{sec:simulation}

Our protocol~$ \crossoverKeProtocol $ (Protocol~\ref{prot:matching_short}) allows for the computation of a maximum matching between a fixed set of patient-donor pairs. However, kidney exchange is dynamic by nature, i.e., patient-donor pairs arrive at and leave from the platform over time. In this section, we analyze the performance of a dynamic kidney exchange platform (cf. Section~\ref{sec:overview}) using our protocol compared to a non-privacy-preserving approach (as explained in Section~\ref{sec:overview}). To this end, we developed a simulation framework based on \text{DESMO-J}~\cite{DESMOJ} which is a discrete event simulation framework written in Java.

Recall that at the core of a kidney exchange platform is the pool of patient-donor pairs who want to find an exchange partner. We simulate the computed transplants over time using our protocol~$ \crossoverKeProtocol $ (Protocol~\ref{prot:matching_short}) for computing the matching inside the pool at a certain point in time. Depending on the number of pairs in the pool at the time of a match run, the runtime of our protocol may be infeasible. This means that the protocol execution does not finish before the next match run is scheduled, e.g., if we compute a matching once a week, the protocol runtime should be below $ 7 $ days. 
In such a case, we split the pool into multiple sub-pools and distribute the pairs uniformly at random among these. Of course, this may lead to fewer pairs being matched in the privacy-preserving case than in the conventional case (where a matching can always be computed efficiently among all pairs in the pool). Our policy is to split the pool into the minimum number of equally sized sub-pools whose execution time is below the match run interval. 

The goal of our simulations is then to evaluate the impact of this pool splitting on the number of patient-donor pairs that can receive a transplant. Thereby, we can evaluate the impact of the larger runtimes of our privacy-preserving protocol compared to the non-privacy-preserving solution and thus determine the practicality of our protocol when used in a real-world kidney exchange platform.

\subsection{Data Set}

We use a data set from the Organ Procurement and Transplantation Network (OPTN) provided by the United Network for Organ Sharing (UNOS) containing all patients and donors that have participated in their living donor exchange program between 27th October 2010 and 29th December 2020.~\footnote{The data reported here have been supplied by the United Network for Organ Sharing as the contractor for the Organ Procurement and Transplantation Network. The interpretation and reporting of these data are the responsibility of the author(s) and in no way should be seen as an official policy of or interpretation by the OPTN or the U.S.\ Government.} The data set contains 2738 patients that were registered with UNOS. There are 150 patients that have registered with more than one donor over time. We consider these as different unique pairs leading to an overall number of 2913 unique patient-donor pairs from which we sample in our simulations. We use the straight-forward approach of choosing the entering patient-donor pairs as described in~\cite{AshlagiKidneyExchangeOperations2021}, i.e., we sample randomly from all pairs when inserting a new pair into the pool.

\subsection{Simulation Setup}

For each simulation run, we consider a time horizon of five years and evaluate the number of patients that receive a kidney transplant as well as the average waiting time (i.e., the average time a patient has to wait until receiving a transplant). We compare the approach using our protocol~$ \crossoverKeProtocol $ (Protocol~\ref{prot:matching_short}) to a conventional approach which is not resistant against manipulation and does not protect the privacy of the patient-donor pairs' medical data. For the conventional approach, we assume that computing the matching is done instantly using a state-of-the-art matching algorithm. In the privacy-preserving case, we simulate the protocol execution using a non-privacy-preserving implementation of our protocol and then schedule the completion of the protocol execution by adding the runtime measured in Section~\ref{sub:runtime_measurements} to the simulation time. Thereby, we reduce the time and resources required by the simulations which allows us to simulate larger parameter ranges.  

A platform as described in Section~\ref{sec:overview} suggests the following parameters that can vary among different platforms (and countries) and for which we run simulations to determine those platforms (and countries) for which our privacy-preserving protocol scales~best:

The \emph{arrival rate} states the frequency at which new patient-donor pairs register with the platform (e.g., arrival rate $ 2 $ indicates that a new pair arrives every $ 2 $ days). In the UNOS data set, a new pair registers every $ 1.4 $ days on average. However, the arrival rate differs among existing platforms depending mainly on the size of the population that participates in the exchange. Therefore, we follow Ashlagi et al.~\cite{AshlagiFrequencies2018} who suggest arrival rates between $ 1 $ and~$ 14 $~days.  
	
The \emph{match run interval} is the interval at which a matching among the pairs in the pool is computed (e.g., a match run interval of size~$ 1 $ indicates that a matching is computed on a daily basis). While platforms in the US such as APKD (Alliance for Paired Kidney Donation) and NKR (National Kidney Registry) have switched to daily match runs or at least multiple match runs per week (UNOS), other countries such as Australia, UK, Netherlands, and Canada compute a matching only once every 2-4 months~\cite{AshlagiKidneyExchangeOperations2021,Biro_EuropeanKE_2019,Cole_CanadianExchange_2015,Ferrari_KidneyPairedDonation_2015}. Therefore, we consider values spanning from one day to four months.
	
The \emph{departure rate} is the probability that a patient-donor pair leaves the pool due to illness/death of donor/patient or due to~a~donor backing out. Based on data from APKD, Ashlagi et al.~\cite{AshlagiFrequencies2018} estimate that a pair on average stays for about $ 420 $ days in the pool without being matched. They simulate this by assuming that a pair leaves each day with probability~$ \frac{1}{420} $. Besides, they also consider pairs staying for $ 800 $ and $ 1000 $ days on average. We follow their approach and consider pairs staying for $ 400 $, $ 800 $, and $ 1000 $ days on average.

The \emph{match refusal probability} is the probability that a match is refused by the patient-donor pair or the hospital for any other reason than a positive crossmatch.\footnote{A crossmatch is a test stating if a patient and a donor are medically compatible. In Protocol~\ref{prot:matching_short}, we compute a virtual crossmatch which does not always match the result of a physical crossmatch. A physical crossmatch, however, can only be carried out after a match is computed as it requires mixing the blood of patient and donor.} For large US exchanges, this probability lies between $ 20\% $ and $ 35\% $~\cite{AshlagiFrequencies2018,AgrarwalKidneyExchangeProductivity2018} whereas in reports on kidney exchange platforms in Europe there is no mention of such high match refusal probabilities~\cite{Biro_EuropeanKE_2019,Ferrari_KidneyPairedDonation_2015}. 
As we aim to cover all existing platforms, we consider match refusal probabilities $ 0\% $, $ 10\% $, $ 20\% $, $ 30\% $, and $ 40\% $.
	
The \emph{crossmatch failure probability} states the failure rate of offered matches due to a positive crossmatch. Platforms in the US (AKPD, UNOS, NKR)~\cite{AshlagiKidneyExchangeOperations2021} as well as in other countries, e.g., Canada~\cite{Cole_CanadianExchange_2015} report that the rate lies at about $ 35\% $ for highly sensitized patients and $ 10\% $ for others. We adopt these values for our~simulations.
	
The \emph{reentering delay} indicates the time it takes for a patient-donor pair to reenter the pool in case of a match offer and a positive crossmatch failure or a refusal of a matched pair. We follow Ashlagi et al.~\cite{AshlagiFrequencies2018} who suggest a delay of $ 2 $ days for reentering to account for a match refusal and $ 7 $ days to account for a positive crossmatch.

Table~\ref{tab:simulation_parameters} shows all parameters together with the values for which we run simulations. For each combination, we execute $ 50 $ simulation runs with different seeds for the random sampling of entering pairs.

\subsection{Evaluation Results}\label{sub:eval}

The performance of the privacy-preserving approach depends highly on the runtime of the underlying SMPC protocol which in turn depends on the network latency~$ L $ (cf.\ Section~\ref{sub:runtime_measurements}). Therefore, we evaluate the privacy-preserving approach for three different scenarios each using the runtime for a different latency (cf.\ Table~\ref{tab:runtimes}).

\begin{table}[t]
	\caption{Simulation parameters and their respective values.\label{tab:simulation_parameters}}
	\begin{tabular}{l | l}
	parameter									& values																\\ \hline
	arrival rate								& $ 1, 2, 4, 7, 14 $ days													\Tstrut\\
	match run interval							& $ 1, 2, 4, 7, 14, 30, 60, 120 $ days										\\
	average time before departure					& $ 400, 800, 1000 $ days													\\
	match refusal probability					& $ 0, 10, 20, 30, 40 $ $ \% $												\\
										
\end{tabular}
\end{table}

\begin{table*}[t!]
	\caption{Comparison between the number of transplants measured for the conventional approach and the privacy-preserving approach (for different values for the latency $ L $) with a departure rate of $ 1/400 $ and a match refusal probability of $ 20\% $. The percentages state how close the privacy-preserving approach comes to the conventional approach.\label{tab:simulation_comp_ref02}}
	\centering
\small
\begin{tabular}{c | c | c | c | c | c | c | c | c}
\multicolumn{1}{c}{}& \multicolumn{1}{c}{} & \multicolumn{1}{c}{\underline{conv.\ model}}	& \multicolumn{6}{c}{\underline{privacy-preserving model}} 	\\
arrival				& match run	& \multirow{2}{*}{transplants}	& \multicolumn{2}{c}{$ L = 1\textit{ms} $} & \multicolumn{2}{c}{$ L = 5\textit{ms} $} & \multicolumn{2}{c}{$ L = 10\textit{ms} $} \Tstrut\\
rate				& interval	& 				& transplants		& percentage							& transplants		& percentage							& transplants		& percentage\\ 	\hline
\multirow{8}{*}{1}	& 1		& 690.52			& 682.24			& \phantom{0}\cellcolor{green!25}98.80\%& 676.52			& \phantom{0}\cellcolor{green!25}97.97\%& 676.80			& \phantom{0}\cellcolor{green!25}98.01\%	\Tstrut\\ 
					& 2		& 689.80			& 672.88			& \phantom{0}\cellcolor{green!25}97.55\%& 669.24			& \phantom{0}\cellcolor{green!25}97.02\%& 661.08			& \phantom{0}\cellcolor{green!25}95.84\%	\\ 
					& 4		& 683.72			& 655.60			& \phantom{0}\cellcolor{green!25}95.89\%& 645.16			& \phantom{0}94.36\%					& 640.96			& \phantom{0}93.75\%						\\
					& 7		& 678.84			& 642.36			& \phantom{0}94.63\%					& 621.04			& \phantom{0}91.49\%					& 615.64			& \phantom{0}90.69\%						\\
					& 14	& 658.92			& 604.04			& \phantom{0}91.67\%					& 578.36			& \phantom{0}87.77\%					& 560.68			& \phantom{0}85.09\%						\\
					& 30	& 612.24			& 529.36			& \phantom{0}86.46\%					& 489.52			& \phantom{0}79.96\%					& 480.20			& \phantom{0}78.43\%						\\
					& 60	& 555.12			& 440.20			& \phantom{0}79.30\%					& 399.44			& \phantom{0}71.96\%					& 383.88			& \phantom{0}69.15\%						\\
					& 120	& 460.48			& 329.84			& \phantom{0}71.63\%					& 297.04			& \phantom{0}64.51\%					& 282.84			& \phantom{0}61.42\%						\\ \hline
\multirow{8}{*}{2}	& 1		& 298.96			& 299.12			& \cellcolor{green!25}100.05\%			& 295.52			& \phantom{0}\cellcolor{green!25}98.85\%& 295.28			& \phantom{0}\cellcolor{green!25}98.77\%	\Tstrut\\ 
					& 2		& 298.68			& 295.12			& \phantom{0}\cellcolor{green!25}98.81\%& 294.60			& \phantom{0}\cellcolor{green!25}98.63\%& 290.28			& \phantom{0}\cellcolor{green!25}97.19\%	\\ 
					& 4		& 296.88			& 288.68			& \phantom{0}\cellcolor{green!25}97.24\%& 283.44			& \phantom{0}\cellcolor{green!25}95.47\%& 286.52			& \phantom{0}\cellcolor{green!25}96.51\%	\\
					& 7		& 291.36			& 283.72			& \phantom{0}\cellcolor{green!25}97.38\%& 279.44			& \phantom{0}\cellcolor{green!25}95.91\%& 277.44			& \phantom{0}\cellcolor{green!25}95.22\%	\\
					& 14	& 282.64			& 272.80			& \phantom{0}\cellcolor{green!25}96.52\%& 264.88			& \phantom{0}93.72\%					& 260.08			& \phantom{0}92.02\%						\\
					& 30	& 264.16			& 242.52			& \phantom{0}91.81\%					& 229.44			& \phantom{0}86.86\%					& 227.56			& \phantom{0}86.14\%						\\
					& 60	& 239.00			& 206.52			& \phantom{0}86.41\%					& 188.56			& \phantom{0}78.90\%					& 182.56			& \phantom{0}76.38\%						\\
					& 120	& 199.96			& 158.92			& \phantom{0}79.48\%					& 146.52			& \phantom{0}73.27\%					& 140.52			& \phantom{0}70.27\%						\\ \hline
\multirow{8}{*}{4}	& 1		& 123.48			& 122.44			& \phantom{0}\cellcolor{green!25}99.16\%& 122.48			& \phantom{0}\cellcolor{green!25}99.19\%& 124.88			& \cellcolor{green!25}101.13\%				\Tstrut\\ 
					& 2		& 123.20			& 122.40			& \phantom{0}\cellcolor{green!25}99.35\%& 124.40			& \cellcolor{green!25}100.97\%			& 123.20			& \cellcolor{green!25}100.00\%				\\ 
					& 4		& 123.84			& 122.28			& \phantom{0}\cellcolor{green!25}98.74\%& 122.08			& \phantom{0}\cellcolor{green!25}98.58\%& 119.00			& \phantom{0}\cellcolor{green!25}96.09\%	\\
					& 7		& 122.20			& 119.64			& \phantom{0}\cellcolor{green!25}97.91\%& 120.04			& \phantom{0}\cellcolor{green!25}98.23\%& 117.64			& \phantom{0}\cellcolor{green!25}96.27\%	\\
					& 14	& 117.44			& 115.00			& \phantom{0}\cellcolor{green!25}97.92\%& 113.84			& \phantom{0}\cellcolor{green!25}96.93\%& 113.08			& \phantom{0}\cellcolor{green!25}96.29\%	\\
					& 30	& 109.80			& 104.12			& \phantom{0}94.83\%					& 103.04			& \phantom{0}93.84\%					& 101.20			& \phantom{0}92.17\%						\\
					& 60	& \phantom{0}99.56	& \phantom{0}93.20	& \phantom{0}93.61\%					& \phantom{0}87.92	& \phantom{0}88.31\%					& \phantom{0}85.64	& \phantom{0}86.02\%						\\
					& 120	& \phantom{0}80.72	& \phantom{0}70.76	& \phantom{0}87.66\%					& \phantom{0}67.32	& \phantom{0}83.40\%					& \phantom{0}66.56	& \phantom{0}82.46\%						\\ \hline
\multirow{8}{*}{7}	& 1		& \phantom{0}56.52	& \phantom{0}57.16	& \cellcolor{green!25}101.13\%			& \phantom{0}57.08	& \cellcolor{green!25}100.99\%			& \phantom{0}57.00	& \cellcolor{green!25}100.85\%				\Tstrut\\ 
					& 2		& \phantom{0}56.40	& \phantom{0}56.56	& \cellcolor{green!25}100.28\%			& \phantom{0}57.28	& \cellcolor{green!25}101.56\%			& \phantom{0}56.12	& \phantom{0}\cellcolor{green!25}99.50\%	\\ 
					& 4		& \phantom{0}55.68	& \phantom{0}56.40	& \cellcolor{green!25}101.29\%			& \phantom{0}57.44	& \cellcolor{green!25}103.16\%			& \phantom{0}55.00	& \phantom{0}\cellcolor{green!25}98.78\%	\\
					& 7		& \phantom{0}56.92	& \phantom{0}56.60	& \phantom{0}\cellcolor{green!25}99.44\%& \phantom{0}55.48	& \phantom{0}\cellcolor{green!25}97.47\%& \phantom{0}55.84	& \phantom{0}\cellcolor{green!25}98.10\%	\\
					& 14	& \phantom{0}54.20	& \phantom{0}54.20	& \cellcolor{green!25}100.00\%			& \phantom{0}53.92	& \phantom{0}\cellcolor{green!25}99.48\%& \phantom{0}52.76	& \phantom{0}\cellcolor{green!25}97.34\%	\\
					& 30	& \phantom{0}50.80	& \phantom{0}50.84	& \cellcolor{green!25}100.08\%			& \phantom{0}49.00	& \phantom{0}\cellcolor{green!25}96.46\%& \phantom{0}48.32	& \phantom{0}\cellcolor{green!25}95.12\%	\\
					& 60	& \phantom{0}45.68	& \phantom{0}44.88	& \phantom{0}\cellcolor{green!25}98.25\%& \phantom{0}42.72	& \phantom{0}93.52\%					& \phantom{0}40.64	& \phantom{0}88.97\%						\\
					& 120	& \phantom{0}38.16	& \phantom{0}38.12	& \phantom{0}\cellcolor{green!25}99.90\%& \phantom{0}33.28	& \phantom{0}87.21\%					& \phantom{0}31.28	& \phantom{0}81.97\%						\\ \hline
\multirow{8}{*}{14}	& 1		& \phantom{0}20.00	& \phantom{0}20.76	& \cellcolor{green!25}103.80\%			& \phantom{0}19.76	& \phantom{0}\cellcolor{green!25}98.80\%& \phantom{0}19.60	& \phantom{0}\cellcolor{green!25}98.00\%	\Tstrut\\ 
					& 2		& \phantom{0}19.76	& \phantom{0}19.48	& \phantom{0}\cellcolor{green!25}98.58\%& \phantom{0}20.00	& \cellcolor{green!25}101.21\%			& \phantom{0}20.12	& \phantom{0}\cellcolor{green!25}101.82\%	\\ 
					& 4		& \phantom{0}19.96	& \phantom{0}21.04	& \cellcolor{green!25}105.41\%			& \phantom{0}21.12	& \cellcolor{green!25}105.81\%			& \phantom{0}21.72	& \cellcolor{green!25}108.82\%				\\
					& 7		& \phantom{0}19.96	& \phantom{0}20.00	& \cellcolor{green!25}100.20\%			& \phantom{0}20.00	& \cellcolor{green!25}100.20\%			& \phantom{0}19.88	& \phantom{0}\cellcolor{green!25}99.60\%	\\
					& 14	& \phantom{0}20.44	& \phantom{0}20.32	& \phantom{0}\cellcolor{green!25}99.41\%& \phantom{0}20.32	& \phantom{0}\cellcolor{green!25}99.41\%& \phantom{0}20.24	& \phantom{0}\cellcolor{green!25}99.02\%	\\
					& 30	& \phantom{0}17.48	& \phantom{0}17.48	& \cellcolor{green!25}100.00\%			& \phantom{0}17.40	& \phantom{0}\cellcolor{green!25}99.54\%& \phantom{0}17.44	& \phantom{0}\cellcolor{green!25}99.77\%	\\
					& 60	& \phantom{0}17.08	& \phantom{0}17.08	& \cellcolor{green!25}100.00\%			& \phantom{0}17.12	& \cellcolor{green!25}100.23\%			& \phantom{0}17.04	& \phantom{0}\cellcolor{green!25}99.77\%	\\
					& 120	& \phantom{0}13.72	& \phantom{0}13.72	& \cellcolor{green!25}100.00\%			& \phantom{0}13.76	& \cellcolor{green!25}100.29\%			& \phantom{0}13.24	& \phantom{0}\cellcolor{green!25}96.50\%	\\ 
\end{tabular}
\end{table*}

As we cannot present the results for all parameter values in this paper, we focus on an average time before departure of $ 400 $ days and a match refusal probability of $ 20\% $ as these are the values most commonly found in existing platforms. Besides, our simulations have shown that departure rate and match refusal probability only slightly influence the performance of the privacy-preserving approach compared to the conventional approach. 
Results for additional values are included in the Appendix~\ref{app:sim}.

Table~\ref{tab:simulation_comp_ref02} shows the number of transplants for a departure rate of~$ 1/400 $ and a match refusal probability of~$ 20\% $.
To compare the two approaches, we compute the percentage of transplants measured for the privacy-preserving approach compared to those measured for the conventional approach (e.g., a percentage of $ 95\% $ indicates that the privacy-preserving approach only leads to $ 5\% $ fewer transplants over the considered time horizon of five years). We highlight all entries where the percentage is larger than $ 95\% $ in Table~\ref{tab:simulation_comp_ref02} indicating those parameters for which the negative impact of the privacy-preserving approach on the number of transplants is very small. 

\paragraph{\textbf{General Observations}}

The privacy-preserving approach is better suited for large arrival rates. This was to be expected since in the privacy-preserving approach the pool has to be split into multiple sub-pools if the number of pairs inside the pool becomes too large. This potentially leads to two pairs which are compatible ending up in different sub-pools. Thus, it is better for the privacy-preserving approach if the patient-donor pairs arrive less frequently as this also means that on average there are fewer patient-donor pairs in the pool which in turn decreases the necessity for pool splitting. 

The privacy-preserving approach performs best if the match run interval is low. This is also as expected since a low match run interval also means that in general fewer patient-donor pairs are in the pool at each match run. Furthermore, if match runs are executed frequently, the probability that two potentially matching pairs are in the same sub-pool is higher since the pairs are distributed uniformly at random among all sub-pools for each match run. 

The privacy-preserving approach sometimes even outperforms the conventional approach. We prescribe this to the fact that there is no known best policy for choosing the match run interval~\cite{AshlagiFrequencies2018}. While, in general, executing match runs with a high frequency seems to be best, there are cases where waiting for further pairs to enter the pool can be better. Consider the following example: Pair~$ A $ is compatible with pairs~$ B $ and $ C $ whereas pair~$ D $ is only compatible with pair~$ B $. If a match run is executed before pairs~$ C $ and $ D $ arrive, only $ A $ and $ B $ can be matched. If one however waits until pairs $ C $ and $ D $ have arrived, then $ A $ can be matched with $ C $ and $ B $ with $ D $. 

\paragraph{\textbf{Influence of the Latency}}
We observe that the smaller the latency in the network connecting the three computing peers, the better the performance of the privacy-preserving approach. This was to be expected as the runtime of the protocol determines the number of pairs for which the pool can still be executed without splitting it into sub-pools. For latency $ L = 1\textit{ms} $, we are able to find nearly the same number of transplants as in the conventional approach independent of the arrival rate when considering a small match run interval (e.g., over the complete time span of $ 5 $ years there are on average only $ 8.28 $ fewer transplants for arrival rate $ 1 $ and even $ 0.76 $ additional transplants for arrival rate $ 14 $). We observe similar findings for $ L = 5\textit{ms} $ while the percentages are in general a bit smaller than for $ L = 1\textit{ms} $. The same holds for $ L = 10\textit{ms} $. However, even for such a large latency, the number of transplants for small match run intervals is still very close to the conventional approach (e.g., independent of the arrival rate the privacy-preserving approach still achieves more than $ 98\% $ of the number of transplants for the conventional approach if a match runs are executed daily).

\paragraph{\textbf{Average Waiting Time}}
The average waiting time (i.e., the average time a patient waits until receiving a transplant) is nearly always larger for the privacy-preserving approach as the SMPC protocol execution may take up to $ 7 $ days whereas the runtime of the matching algorithm in the conventional approach is negligible. However, for those parameters where the number of transplants in the privacy-preserving approach is close to the conventional approach, the average waiting time only differs by a few days. This is acceptable as even in the conventional approach the patients on average wait multiple months before receiving a transplant. 
The~corresponding simulation results can be found in Appendix~\ref{app:waiting_time}.

\paragraph{\textbf{Summary}}
The negative impact of the privacy-preserving approach can be considered negligible for arrival rates of $ 4 $ to $ 14 $ days if the match run interval is between $ 1 $ and $ 7 $ days and for arrival rates of $ 1 $ or $ 2 $ days if the match run interval equals $ 1 $ or $ 2 $ days. Thus, our approach is even practical for large platforms such as UNOS in the US where patient-donor pairs arrive very frequently. Also, our approach performs best for those parameters for which~the~number of transplants is highest. This coincides with those values which one would prefer in real-world platforms. Thus, the worse performance of our approach for large match run intervals is not of relevance in practice. Finally, while $ 95\% $ of the number of transplants compared to the conventional approach may not seem acceptable in countries where kidney exchange is already practiced, our approach would increase the number of possible transplants from $ 0\% $ to $ 95\% $ of the maximum possible number of transplants in countries such as Germany where kidney exchange is still not practiced.

\section{Conclusion and Future Work}\label{sec:conclusion}

We have presented a privacy-preserving protocol for crossover kidney exchange and evaluated its performance for different properties of the underlying network. Our protocol clearly outperforms the existing protocol for kidney exchange and offers reasonable runtimes for up to $ 64 $ patient-donor pairs depending on the network latency. We evaluated the practicality of our protocol when used in a dynamic kidney exchange platform and compared it to a conventional approach. Overall, the cost induced by adding privacy is low. Specifically, if match runs are executed frequently, the difference between the privacy-preserving and the conventional approach is negligible w.r.t.\ the number of transplants that can~be~found. 

One direction for future work is the development of protocols that can handle larger exchange cycles or chains such that also platforms that allow for such more sophisticated exchange structures can be made privacy-preserving. Furthermore, we will strive to add the possibility of prioritization and robustness (in case a computed match fails) to our protocol further closing the gap between conventional and privacy-preserving kidney exchange.

\begin{acks}
	This work was funded by the Deutsche Forschungsgemeinschaft (DFG, German Resarch Foundation) - project number (419340256) and NSF grant CCF-1646999. Any opinion, findings, and conclusions or recommendations expressed in this material are those of the author(s) and do not necessarily reflect the views of the National Science Foundation.
	Computations for our simulations were performed with computing resources granted by RWTH Aachen University under Project RWTH0438.
\end{acks}

\bibliographystyle{ACM-Reference-Format}
\bibliography{references}


\begin{thebibliography}{40}


\ifx \showCODEN    \undefined \def \showCODEN     #1{\unskip}     \fi
\ifx \showDOI      \undefined \def \showDOI       #1{#1}\fi
\ifx \showISBNx    \undefined \def \showISBNx     #1{\unskip}     \fi
\ifx \showISBNxiii \undefined \def \showISBNxiii  #1{\unskip}     \fi
\ifx \showISSN     \undefined \def \showISSN      #1{\unskip}     \fi
\ifx \showLCCN     \undefined \def \showLCCN      #1{\unskip}     \fi
\ifx \shownote     \undefined \def \shownote      #1{#1}          \fi
\ifx \showarticletitle \undefined \def \showarticletitle #1{#1}   \fi
\ifx \showURL      \undefined \def \showURL       {\relax}        \fi
\providecommand\bibfield[2]{#2}
\providecommand\bibinfo[2]{#2}
\providecommand\natexlab[1]{#1}
\providecommand\showeprint[2][]{arXiv:#2}

\bibitem[Abraham et~al\mbox{.}(2007)]%
        {AbrahamClearingAlgorithms2007}
\bibfield{author}{\bibinfo{person}{David~J Abraham}, \bibinfo{person}{Avrim
  Blum}, {and} \bibinfo{person}{Tuomas Sandholm}.}
  \bibinfo{year}{2007}\natexlab{}.
\newblock \showarticletitle{Clearing Algorithms for Barter Exchange Markets:
  Enabling Nationwide Kidney Exchange}.
\newblock \bibinfo{journal}{\emph{ACM Conference on Electronic Commerce}}.
\newblock


\bibitem[Agarwal et~al\mbox{.}(2018)]%
        {AgrarwalKidneyExchangeProductivity2018}
\bibfield{author}{\bibinfo{person}{Nikhil Agarwal}, \bibinfo{person}{Itai
  Ashlagi}, \bibinfo{person}{Eduardo Azevedo}, \bibinfo{person}{Clayton
  Featherstone}, {and} \bibinfo{person}{{\"O}mer Karaduman}.}
  \bibinfo{year}{2018}\natexlab{}.
\newblock \showarticletitle{What Matters for the Productivity of Kidney
  Exchange?}. In \bibinfo{booktitle}{\emph{AEA Papers and Proceedings}},
  Vol.~\bibinfo{volume}{108}.
\newblock


\bibitem[Aly et~al\mbox{.}(2013)]%
        {Aly_Bipartite_2013}
\bibfield{author}{\bibinfo{person}{Abdelrahaman Aly}, \bibinfo{person}{Edouard
  Cuvelier}, \bibinfo{person}{Sophie Mawet}, \bibinfo{person}{Olivier Pereira},
  {and} \bibinfo{person}{Mathieu Van~Vyve}.} \bibinfo{year}{2013}\natexlab{}.
\newblock \showarticletitle{Securely solving simple combinatorial graph
  problems}.
\newblock \bibinfo{journal}{\emph{International Conference on Financial
  Cryptography and Data Security}}.
\newblock


\bibitem[Anandan and Clifton(2017)]%
        {Anandan_Bipartite_2017}
\bibfield{author}{\bibinfo{person}{Balamurugan Anandan} {and}
  \bibinfo{person}{Chris Clifton}.} \bibinfo{year}{2017}\natexlab{}.
\newblock \showarticletitle{Secure minimum weighted bipartite matching}.
\newblock \bibinfo{journal}{\emph{IEEE Conference on Dependable and Secure
  Computing}}.
\newblock


\bibitem[Andersson and Kratz(2020)]%
        {Anderson_PairwiseSweden_2020}
\bibfield{author}{\bibinfo{person}{Tommy Andersson} {and}
  \bibinfo{person}{J{\"o}rgen Kratz}.} \bibinfo{year}{2020}\natexlab{}.
\newblock \showarticletitle{Pairwise kidney exchange over the blood group
  barrier}.
\newblock \bibinfo{journal}{\emph{The Review of Economic Studies}}
  \bibinfo{volume}{87}, \bibinfo{number}{3}.
\newblock


\bibitem[Ashlagi et~al\mbox{.}(2018)]%
        {AshlagiFrequencies2018}
\bibfield{author}{\bibinfo{person}{Itai Ashlagi}, \bibinfo{person}{Adam
  Bingaman}, \bibinfo{person}{Maximilien Burq}, \bibinfo{person}{Vahideh
  Manshadi}, \bibinfo{person}{David Gamarnik}, \bibinfo{person}{Cathi Murphey},
  \bibinfo{person}{Alvin~E Roth}, \bibinfo{person}{Marc~L Melcher}, {and}
  \bibinfo{person}{Michael~A Rees}.} \bibinfo{year}{2018}\natexlab{}.
\newblock \showarticletitle{Effect of match-run frequencies on the number of
  transplants and waiting times in kidney exchange}.
\newblock \bibinfo{journal}{\emph{American Journal of Transplantation}}
  \bibinfo{volume}{18}, \bibinfo{number}{5} (\bibinfo{year}{2018}).
\newblock


\bibitem[Ashlagi and Roth(2021)]%
        {AshlagiKidneyExchangeOperations2021}
\bibfield{author}{\bibinfo{person}{Itai Ashlagi} {and} \bibinfo{person}{Alvin~E
  Roth}.} \bibinfo{year}{2021}\natexlab{}.
\newblock \bibinfo{booktitle}{\emph{Kidney Exchange: An Operations
  Perspective}}.
\newblock \bibinfo{type}{{T}echnical {R}eport}. \bibinfo{institution}{National
  Bureau of Economic Research}.
\newblock


\bibitem[Ben-Or et~al\mbox{.}(1988)]%
        {BenOr_ConstantRoundsMult_1988}
\bibfield{author}{\bibinfo{person}{Michael Ben-Or}, \bibinfo{person}{Shafi
  Goldwasser}, {and} \bibinfo{person}{Avi Wigderson}.}
  \bibinfo{year}{1988}\natexlab{}.
\newblock \showarticletitle{Completeness theorems for non-cryptographic
  fault-tolerant distributed computations}. In \bibinfo{booktitle}{\emph{ACM
  Symposium on Theory of Computing}}. \bibinfo{publisher}{ACM}.
\newblock


\bibitem[Berge(1957)]%
        {BergeTheorem1957}
\bibfield{author}{\bibinfo{person}{Claude Berge}.}
  \bibinfo{year}{1957}\natexlab{}.
\newblock \showarticletitle{Two Theorems in Graph Theory}.
\newblock \bibinfo{journal}{\emph{Proceedings of the National Academy of
  Sciences of the United States of America}} \bibinfo{volume}{43},
  \bibinfo{number}{9} (\bibinfo{year}{1957}).
\newblock


\bibitem[Bir{\'o} et~al\mbox{.}(2019a)]%
        {Biro_EuropeanKE_2019}
\bibfield{author}{\bibinfo{person}{P{\'e}ter Bir{\'o}},
  \bibinfo{person}{Bernadette Haase-Kromwijk}, \bibinfo{person}{Tommy
  Andersson}, \bibinfo{person}{Eyj{\'o}lfur~Ingi {\'A}sgeirsson},
  \bibinfo{person}{Tatiana Baltesov{\'a}}, \bibinfo{person}{Ioannis Boletis},
  \bibinfo{person}{Catarina Bolotinha}, \bibinfo{person}{Gregor Bond},
  \bibinfo{person}{Georg B{\"o}hmig}, \bibinfo{person}{Lisa Burnapp},
  {et~al\mbox{.}}} \bibinfo{year}{2019}\natexlab{a}.
\newblock \showarticletitle{Building Kidney Exchange Programmes in Europe -- An
  Overview of Exchange Practice and Activities}.
\newblock \bibinfo{journal}{\emph{Transplantation}} \bibinfo{volume}{103},
  \bibinfo{number}{7} (\bibinfo{year}{2019}).
\newblock


\bibitem[Bir{\'o} et~al\mbox{.}(2019b)]%
        {Biro_EuropeanModellingKE_2019}
\bibfield{author}{\bibinfo{person}{P{\'e}ter Bir{\'o}}, \bibinfo{person}{Joris
  van~de Klundert}, \bibinfo{person}{David Manlove}, \bibinfo{person}{William
  Pettersson}, \bibinfo{person}{Tommy Andersson}, \bibinfo{person}{Lisa
  Burnapp}, \bibinfo{person}{Pavel Chromy}, \bibinfo{person}{Pablo Delgado},
  \bibinfo{person}{Piotr Dworczak}, \bibinfo{person}{Bernadette Haase},
  {et~al\mbox{.}}} \bibinfo{year}{2019}\natexlab{b}.
\newblock \showarticletitle{Modelling and optimisation in european kidney
  exchange programmes}. In \bibinfo{booktitle}{\emph{European Journal of
  Operational Research}}. \bibinfo{publisher}{Elsevier}.
\newblock


\bibitem[Blanton and Saraph(2015)]%
        {Blanton_Bipartite_2015}
\bibfield{author}{\bibinfo{person}{Marina Blanton} {and}
  \bibinfo{person}{Siddharth Saraph}.} \bibinfo{year}{2015}\natexlab{}.
\newblock \showarticletitle{Oblivious maximum bipartite matching size algorithm
  with applications to secure fingerprint identification}.
\newblock \bibinfo{journal}{\emph{European Symposium on Research in Computer
  Security}}.
\newblock


\bibitem[Blanton et~al\mbox{.}(2013)]%
        {Blanton_Bipartite_2013}
\bibfield{author}{\bibinfo{person}{Marina Blanton}, \bibinfo{person}{Aaron
  Steele}, {and} \bibinfo{person}{Mehrdad Alisagari}.}
  \bibinfo{year}{2013}\natexlab{}.
\newblock \showarticletitle{Data-oblivious graph algorithms for secure
  computation and outsourcing}.
\newblock \bibinfo{journal}{\emph{ACM SIGSAC symposium on Information, computer
  and communications security}}.
\newblock


\bibitem[Blum(1990)]%
        {Blum_Matching_1990}
\bibfield{author}{\bibinfo{person}{Norbert Blum}.}
  \bibinfo{year}{1990}\natexlab{}.
\newblock \showarticletitle{A new approach to maximum matching in general
  graphs}. In \bibinfo{booktitle}{\emph{International Colloquium on Automata,
  Languages, and Programming}}. Springer.
\newblock


\bibitem[Bogdanov et~al\mbox{.}(2015)]%
        {Bogdanov_Estonian_2015}
\bibfield{author}{\bibinfo{person}{Dan Bogdanov}, \bibinfo{person}{Marko
  J{\~o}emets}, \bibinfo{person}{Sander Siim}, {and} \bibinfo{person}{Meril
  Vaht}.} \bibinfo{year}{2015}\natexlab{}.
\newblock \showarticletitle{How the estonian tax and customs board evaluated a
  tax fraud detection system based on secure multi-party computation}. In
  \bibinfo{booktitle}{\emph{International conference on financial cryptography
  and data security}}. Springer.
\newblock


\bibitem[Bogdanov et~al\mbox{.}(2008)]%
        {Bogdanov_Sharemind_2008}
\bibfield{author}{\bibinfo{person}{Dan Bogdanov}, \bibinfo{person}{Sven Laur},
  {and} \bibinfo{person}{Jan Willemson}.} \bibinfo{year}{2008}\natexlab{}.
\newblock \showarticletitle{Sharemind: A framework for fast privacy-preserving
  computations}. In \bibinfo{booktitle}{\emph{European Symposium on Research in
  Computer Security}}. Springer.
\newblock


\bibitem[Breuer et~al\mbox{.}(Code)]%
        {Breuer_Code_2022}
\bibfield{author}{\bibinfo{person}{Malte Breuer}, \bibinfo{person}{Ulrike
  Meyer}, {and} \bibinfo{person}{Susanne Wetzel}.} \bibinfo{year}{Source
  Code}\natexlab{}.
\newblock
  \bibinfo{title}{\url{https://gitlab.com/rwth-itsec/privacy-preserving-crossover-exchange}}.
\newblock
\newblock


\bibitem[Breuer et~al\mbox{.}(2020)]%
        {Breuer_KEprotocol_2020}
\bibfield{author}{\bibinfo{person}{Malte Breuer}, \bibinfo{person}{Ulrike
  Meyer}, \bibinfo{person}{Susanne Wetzel}, {and} \bibinfo{person}{Anja
  M{\"u}hlfeld}.} \bibinfo{year}{2020}\natexlab{}.
\newblock \showarticletitle{A Privacy-Preserving Protocol for the Kidney
  Exchange Problem}. In \bibinfo{booktitle}{\emph{Workshop on Privacy in the
  Electronic Society}}. \bibinfo{publisher}{ACM}.
\newblock


\bibitem[Catrina and De~Hoogh(2010)]%
        {Catrina_PrimitivesSMPC_2010}
\bibfield{author}{\bibinfo{person}{Octavian Catrina} {and}
  \bibinfo{person}{Sebastiaan De~Hoogh}.} \bibinfo{year}{2010}\natexlab{}.
\newblock \showarticletitle{Improved primitives for secure multiparty integer
  computation}.
\newblock \bibinfo{journal}{\emph{International Conference on Security and
  Cryptography for Networks}}.
\newblock


\bibitem[Cole et~al\mbox{.}(2015)]%
        {Cole_CanadianExchange_2015}
\bibfield{author}{\bibinfo{person}{Edward~H Cole}, \bibinfo{person}{Peter
  Nickerson}, \bibinfo{person}{Patricia Campbell}, \bibinfo{person}{Kathy
  Yetzer}, \bibinfo{person}{Nick Lahaie}, \bibinfo{person}{Jeffery Zaltzman},
  {and} \bibinfo{person}{John~S Gill}.} \bibinfo{year}{2015}\natexlab{}.
\newblock \showarticletitle{The Canadian kidney paired donation program: a
  national program to increase living donor transplantation}.
\newblock \bibinfo{journal}{\emph{Transplantation}} \bibinfo{volume}{99},
  \bibinfo{number}{5} (\bibinfo{year}{2015}).
\newblock


\bibitem[{DESMO-J}(enet)]%
        {DESMOJ}
\bibfield{author}{\bibinfo{person}{{DESMO-J}}.}
  \bibinfo{year}{\url{http://desmoj.sourceforge.net}}\natexlab{}.
\newblock \bibinfo{title}{Accessed 30-Sep-2021}.
\newblock
\newblock


\bibitem[Dickerson et~al\mbox{.}(2019)]%
        {Dickerson_FailureAwareKE_2019}
\bibfield{author}{\bibinfo{person}{John~P. Dickerson},
  \bibinfo{person}{Ariel~D. Procaccia}, {and} \bibinfo{person}{Tuomas
  Sandholm}.} \bibinfo{year}{2019}\natexlab{}.
\newblock \showarticletitle{Failure-Aware Kidney Exchange}.
\newblock \bibinfo{journal}{\emph{Management Science}} \bibinfo{volume}{65},
  \bibinfo{number}{4} (\bibinfo{year}{2019}).
\newblock


\bibitem[Doerner et~al\mbox{.}(2016)]%
        {Doerner_Bipartite_2016}
\bibfield{author}{\bibinfo{person}{Jack Doerner}, \bibinfo{person}{David
  Evans}, {and} \bibinfo{person}{Abhi Shelat}.}
  \bibinfo{year}{2016}\natexlab{}.
\newblock \showarticletitle{Secure stable matching at scale}.
\newblock \bibinfo{journal}{\emph{ACM SIGSAC Conference on Computer and
  Communications Security}}.
\newblock


\bibitem[Edmonds(1965)]%
        {EdmondsBlossomAlgorithm1965}
\bibfield{author}{\bibinfo{person}{Jack Edmonds}.}
  \bibinfo{year}{1965}\natexlab{}.
\newblock \showarticletitle{Paths, Trees, and Flowers}.
\newblock \bibinfo{journal}{\emph{Canadian Journal of mathematics}}
  \bibinfo{volume}{17} (\bibinfo{year}{1965}).
\newblock


\bibitem[Ferrari et~al\mbox{.}(2015)]%
        {Ferrari_KidneyPairedDonation_2015}
\bibfield{author}{\bibinfo{person}{Paolo Ferrari}, \bibinfo{person}{Willem
  Weimar}, \bibinfo{person}{Rachel~J Johnson}, \bibinfo{person}{Wai~H Lim},
  {and} \bibinfo{person}{Kathryn~J Tinckam}.} \bibinfo{year}{2015}\natexlab{}.
\newblock \showarticletitle{Kidney paired donation: principles, protocols and
  programs}.
\newblock \bibinfo{journal}{\emph{Nephrology Dialysis Transplantation}}
  \bibinfo{volume}{30}, \bibinfo{number}{8} (\bibinfo{year}{2015}).
\newblock


\bibitem[Gabow(1976)]%
        {GabowMatching1976}
\bibfield{author}{\bibinfo{person}{Harold~N Gabow}.}
  \bibinfo{year}{1976}\natexlab{}.
\newblock \showarticletitle{An Efficient Implementation of Edmonds' Algorithm
  for Maximum Matching on Graphs}.
\newblock \bibinfo{journal}{\emph{J. ACM}} \bibinfo{volume}{23},
  \bibinfo{number}{2} (\bibinfo{year}{1976}).
\newblock


\bibitem[Gabow and Tarjan(1991)]%
        {Gabow_FastMatching_1991}
\bibfield{author}{\bibinfo{person}{Harold~N Gabow} {and}
  \bibinfo{person}{Robert~E Tarjan}.} \bibinfo{year}{1991}\natexlab{}.
\newblock \showarticletitle{Faster scaling algorithms for general graph
  matching problems}.
\newblock \bibinfo{journal}{\emph{J. ACM}} \bibinfo{volume}{38},
  \bibinfo{number}{4} (\bibinfo{year}{1991}).
\newblock


\bibitem[Goldreich(2004)]%
        {GoldreichFoundationsTwo2004}
\bibfield{author}{\bibinfo{person}{Oded Goldreich}.}
  \bibinfo{year}{2004}\natexlab{}.
\newblock \bibinfo{booktitle}{\emph{Foundations of Cryptography: Volume 2 -
  Basic Applications}}.
\newblock \bibinfo{publisher}{Cambridge University Press}.
\newblock


\bibitem[Golle(2006)]%
        {Golle_Bipartite_2006}
\bibfield{author}{\bibinfo{person}{Philippe Golle}.}
  \bibinfo{year}{2006}\natexlab{}.
\newblock \showarticletitle{A private stable matching algorithm}.
\newblock \bibinfo{journal}{\emph{International Conference on Financial
  Cryptography and Data Security}}.
\newblock


\bibitem[Keller(2020)]%
        {KellerMPSPDZ2020}
\bibfield{author}{\bibinfo{person}{Marcel Keller}.}
  \bibinfo{year}{2020}\natexlab{}.
\newblock \showarticletitle{{MP-SPDZ}: A Versatile Framework for Multi-Party
  Computation}. In \bibinfo{booktitle}{\emph{Computer and Communications
  Security}}. \bibinfo{publisher}{ACM}.
\newblock


\bibitem[Keller and Scholl(2014)]%
        {Keller_ORAM_2014}
\bibfield{author}{\bibinfo{person}{Marcel Keller} {and} \bibinfo{person}{Peter
  Scholl}.} \bibinfo{year}{2014}\natexlab{}.
\newblock \showarticletitle{Efficient, oblivious data structures for MPC}.
\newblock \bibinfo{journal}{\emph{International Conference on the Theory and
  Application of Cryptology and Information Security}}.
\newblock


\bibitem[Launchbury et~al\mbox{.}(2012)]%
        {Launchbury_Multiplex_2012}
\bibfield{author}{\bibinfo{person}{John Launchbury}, \bibinfo{person}{Iavor~S
  Diatchki}, \bibinfo{person}{Thomas DuBuisson}, {and} \bibinfo{person}{Andy
  Adams-Moran}.} \bibinfo{year}{2012}\natexlab{}.
\newblock \showarticletitle{Efficient lookup-table protocol in secure
  multiparty computation}.
\newblock \bibinfo{journal}{\emph{ACM SIGPLAN international conference on
  Functional programming}}.
\newblock


\bibitem[Micali and Vazirani(1980)]%
        {MicaliMatching1980}
\bibfield{author}{\bibinfo{person}{Silvio Micali} {and}
  \bibinfo{person}{Vijay~V Vazirani}.} \bibinfo{year}{1980}\natexlab{}.
\newblock \showarticletitle{An {O($\sqrt{\vert V \vert} \vert E \vert $)}
  Algorithm for Finding Maximum Matching in General Graphs}. In
  \bibinfo{booktitle}{\emph{Symposium on Foundations of Computer Science}}.
  IEEE.
\newblock


\bibitem[Organization(eath)]%
        {WHO_2019}
\bibfield{author}{\bibinfo{person}{World~Health Organization}.}
  \bibinfo{year}{Top 10 Causes of Death}\natexlab{}.
\newblock
  \bibinfo{title}{\url{https://www.who.int/news-room/fact-sheets/detail/the-top-10-causes-of-death}}.
\newblock
\newblock
\newblock
\shownote{Accessed 30-Sep-2021}.


\bibitem[Pape and Conradt(1980)]%
        {PapeMatching1980}
\bibfield{author}{\bibinfo{person}{U Pape} {and} \bibinfo{person}{D Conradt}.}
  \bibinfo{year}{1980}\natexlab{}.
\newblock \showarticletitle{Maximales Matching in Graphen}.
\newblock \bibinfo{journal}{\emph{Ausgew{\"a}hlte Operations Research Software
  in FORTRAN}} (\bibinfo{year}{1980}).
\newblock


\bibitem[Procurement and Network(data)]%
        {OPTN_2021}
\bibfield{author}{\bibinfo{person}{Organ Procurement} {and}
  \bibinfo{person}{Transplantation Network}.}
  \bibinfo{year}{\url{https://optn.transplant.hrsa.gov/data/view-data-reports/national-data/}}\natexlab{}.
\newblock \bibinfo{booktitle}{\emph{Accessed 30-Sep-2021}}.
\newblock


\bibitem[Riazi et~al\mbox{.}(2017)]%
        {Riazi_Bipartite_2017}
\bibfield{author}{\bibinfo{person}{M~Sadegh Riazi}, \bibinfo{person}{Ebrahim~M
  Songhori}, \bibinfo{person}{Ahmad-Reza Sadeghi}, \bibinfo{person}{Thomas
  Schneider}, {and} \bibinfo{person}{Farinaz Koushanfar}.}
  \bibinfo{year}{2017}\natexlab{}.
\newblock \showarticletitle{Toward Practical Secure Stable Matching.}
\newblock \bibinfo{journal}{\emph{Proc. Priv. Enhancing Technol.}}
  \bibinfo{volume}{2017}, \bibinfo{number}{1}.
\newblock


\bibitem[Shamir(1979)]%
        {ShamirSecretSharing1979}
\bibfield{author}{\bibinfo{person}{Adi Shamir}.}
  \bibinfo{year}{1979}\natexlab{}.
\newblock \showarticletitle{How to Share a Secret}.
\newblock \bibinfo{journal}{\emph{Commun. ACM}} \bibinfo{volume}{22},
  \bibinfo{number}{11} (\bibinfo{year}{1979}).
\newblock


\bibitem[Syslo et~al\mbox{.}(1983)]%
        {SysloMatchingBook1983}
\bibfield{author}{\bibinfo{person}{Maciej Syslo}, \bibinfo{person}{Narsingh
  Deo}, {and} \bibinfo{person}{Janusz~S Kowalik}.}
  \bibinfo{year}{1983}\natexlab{}.
\newblock \bibinfo{booktitle}{\emph{Discrete Optimization Algorithms with
  Pascal Programs}}.
\newblock \bibinfo{publisher}{Prentice Hall}.
\newblock


\bibitem[W\"{u}ller et~al\mbox{.}(2017)]%
        {WuellerHungarianBartering2017}
\bibfield{author}{\bibinfo{person}{Stefan W\"{u}ller}, \bibinfo{person}{Michael
  Vu}, \bibinfo{person}{Ulrike Meyer}, {and} \bibinfo{person}{Susanne Wetzel}.}
  \bibinfo{year}{2017}\natexlab{}.
\newblock \showarticletitle{Using Secure Graph Algorithms for the
  Privacy-Preserving Identification of Optimal Bartering Opportunities}.
\newblock \bibinfo{journal}{\emph{Workshop on Privacy in the Electronic
  Society}}.
\newblock


\end{thebibliography}

\appendix
\section{Additional Runtime Measurements}\label{appendixA}

In Sections~\ref{sub:runtime_measurements} and \ref{sub:eval}, we always assume a bandwidth of $ B = 1\textit{Gbps} $ in the underlying network. We now show that our protocol (Protocol~\ref{prot:matching_short}) is practical even for smaller bandwidth values. 

Table~\ref{tab:runtimes_lat1ms} shows the runtimes for protocol~$ \crossoverKeProtocol $ (Protocol~\ref{prot:matching_short}) for latency $ L = 1\textit{ms} $ and different bandwidths ranging from $ 10\textit{Mbps} $ to $ 1\textit{Gbps} $. We observe that on average the runtime increases by $ 9\% $ for $ B = \textit{Mbps} $, by $ 18\% $ for $ B = 50\textit{Mbps} $, and by $ 102\% $ for $ B = 10\textit{Mbps} $. Thus, the protocol runtime only slightly increases for bandwidths $ 100\textit{Mbps} $ and $ 50\textit{Mbps} $ and only increases significantly for a very small bandwidth of $ 10\textit{Mbps} $. 

The measured runtimes for the different values of the bandwidth show that the bandwidth has much less influence on the protocol runtime than the network latency (cf. Table~\ref{tab:runtimes}). Furthermore, the fact that there is only a slight runtime increase for bandwidths as small as $ 50\textit{Mbps} $ suggests that the performance of a dynamic kidney exchange platform using our protocol with such bandwidths is very close to the performance measured for $ B = 1\textit{Gbps} $ (cf. Section~\ref{sub:eval}).

\section{Additional Simulation Results}\label{app:sim}

In Section~\ref{sub:eval}, we evaluated the performance of our approach for privacy-preserving kidney exchange compared to the conventional approach for different arrival rates and match run intervals. In this section, we additionally evaluate the results for different match refusal probabilities and departure rates. 

Table~\ref{tab:simulation_comp} contains the number of transplants for the conventional approach as well as for the privacy-preserving approach for match refusal probability $ 0\% $ which is the lowest match refusal probability for which we executed simulations. First of all, we notice that the overall number of transplants is larger for match refusal probability $ 0\% $ than for $ 20\% $ (cf.~Table~\ref{tab:simulation_comp_ref02}) independent of the considered approach and latency. The is an expected observation since due to the larger match refusal probability the percentage of computed matches that do not result in a transplant is larger. 

Comparing the performance of the privacy-preserving approach for the two cases, we observe that on average the performance is about $ 0.09\% $ better for match refusal probability $ 0\% $ than for $ 20\% $. Thus, the difference between the results for the two match refusal probabilities can be considered negligible and we can resume that the performance of the privacy-preserving approach does not depend on the match refusal probability.

Table~\ref{tab:simulation_comp_dep800} shows the number of transplants for a departure rate of $ 1/800 $ whereas in Section~\ref{sub:eval} we considered a departure rate of $ 1/400 $. We observe that the number of transplants is larger for departure rate $ 1/800 $ independent of the considered approach. This is very intuitive since for a smaller departure rate, the parties stay longer in the pool on average which increases their probability of being matched. 

With respect to the performance of the privacy-preserving approach, the difference between the percentage of matched parties between departure rates $ 1/400 $ and $ 1/800 $ is very small, i.e., on average the performance decreases by $ 0.77\% $ for departure rate~$ 1/800 $. Thus, we can conclude that the departure rate does not significantly influence the performance of the privacy-preserving approach compared to the conventional approach although it does influence the overall number of transplants in both approaches.

\begin{table}[!t]
	\caption{Runtime and network traffic of protocol~$ \crossoverKeProtocol $ for different numbers of input peers and different bandwidths $ \textit{B} $ for a fixed latency $ \textit{L} = 1\textit{ms} $.\label{tab:runtimes_lat1ms}}
	\centering
\begin{tabular}{c | r r r r}
	\multirow{2}{*}{input peers} 	& \multicolumn{4}{c}{runtime}   															\\ \cline{2-5}
			& $ 1\textnormal{Gbps} $		& $ 100\textnormal{Mbps} $	& $ 50\textnormal{Mbps} $	& $ 10\textnormal{Mbps} $ 	\Tstrut\\ \hline
	5		& 21s							& 22s						& 24s						& 42s						\Tstrut\\  
	10		& 6m							& 6m						& 7m						& 11m						\\  
	15		& 30m							& 33m						& 36m						& 59m						\\ 
	20		& 96m							& 104m						& 112m						& 188m						\\ 
	25		& 4h							& 4h						& 4.5h						& 8h						\\ 
	30		& 8h							& 9h						& 9.5h						& 16h						\\ 
	35		& 15h							& 16h						& 18h						& 31h						\\ 
	40		& 25h							& 28h						& 30h						& 53h						\\ 
\end{tabular}
\end{table}

\begin{table*}[!t]
	\caption{Comparison between the number of transplants measured for the conventional approach and the privacy-preserving approach (for different values for the latency $ L $) with a departure rate of $ 1/400 $ and a match refusal probability of $ 0\% $. The percentages state how close the privacy-preserving approach comes to the conventional approach.\label{tab:simulation_comp}}
	\small
\centering
\begin{tabular}{c | c | c | c | c | c | c | c | c}
\multicolumn{1}{c}{}& \multicolumn{1}{c}{} & \multicolumn{1}{c}{\underline{conv.\ model}}	& \multicolumn{6}{c}{\underline{privacy-preserving model}} 	\\
arrival				& match run		& \multirow{2}{*}{transplants}				& \multicolumn{2}{c}{$ L = 1\textit{ms} $} & \multicolumn{2}{c}{$ L = 10\textit{ms} $} & \multicolumn{2}{c}{$ L = 20\textit{ms} $} \Tstrut\\
rate				& interval	&  				& transplants		& percentage							& transplants		& percentage							& transplants		& percentage\\ \hline
\multirow{4}{*}{1}	& 1		& 725.28			& 710.12			& \phantom{0}\cellcolor{green!25}97.91\%& 705.64			& \phantom{0}\cellcolor{green!25}97.29\%& 702.80			& \phantom{0}\cellcolor{green!25}96.90\%\Tstrut	\\ 
					& 2		& 719.00			& 705.72			& \phantom{0}\cellcolor{green!25}98.15\%& 701.88			& \phantom{0}\cellcolor{green!25}97.62\%& 695.16			& \phantom{0}\cellcolor{green!25}96.68\%\\ 
					& 4		& 716.36			& 690.72			& \phantom{0}\cellcolor{green!25}96.42\%& 677.00			& \phantom{0}94.51\%					& 670.44			& \phantom{0}93.59\%					\\
					& 7		& 710.76			& 675.40			& \phantom{0}\cellcolor{green!25}95.03\%& 655.40			& \phantom{0}92.21\%					& 646.20			& \phantom{0}90.92\%					\\
					& 14	& 689.48			& 632.48			& \phantom{0}91.73\%					& 612.88			& \phantom{0}88.89\%					& 600.04			& \phantom{0}87.03\%					\\
					& 30	& 655.20			& 567.00			& \phantom{0}86.54\%					& 529.48			& \phantom{0}80.81\%					& 522.68			& \phantom{0}79.77\%					\\
					& 60	& 602.92			& 486.76			& \phantom{0}80.73\%					& 442.92			& \phantom{0}73.46\%					& 432.28			& \phantom{0}71.70\%					\\
					& 120	& 519.48			& 380.72			& \phantom{0}73.29\%					& 337.40			& \phantom{0}64.95\%					& 323.48			& \phantom{0}62.27\%					\\ \hline
\multirow{8}{*}{2}	& 1		& 312.80			& 313.40			& \cellcolor{green!25}100.19\%			& 311.00			& \phantom{0}\cellcolor{green!25}99.42\%& 310.16			& \phantom{0}\cellcolor{green!25}99.16\%\Tstrut\\ 
					& 2		& 314.68			& 312.36			& \phantom{0}\cellcolor{green!25}99.26\%& 310.40			& \phantom{0}\cellcolor{green!25}98.64\%& 308.44			& \phantom{0}\cellcolor{green!25}98.02\%\\ 
					& 4		& 314.64			& 309.40			& \phantom{0}\cellcolor{green!25}98.33\%& 302.48			& \phantom{0}\cellcolor{green!25}96.14\%& 307.36			& \phantom{0}\cellcolor{green!25}97.69\%\\
					& 7		& 309.08			& 305.24			& \phantom{0}\cellcolor{green!25}98.76\%& 298.08			& \phantom{0}\cellcolor{green!25}96.44\%& 296.76			& \phantom{0}\cellcolor{green!25}96.01\%\\
					& 14	& 303.88			& 288.04			& \phantom{0}94.79\%					& 279.64			& \phantom{0}92.02\%					& 273.84			& \phantom{0}90.11\%					\\
					& 30	& 285.80			& 259.28			& \phantom{0}90.72\%					& 249.60			& \phantom{0}87.33\%					& 245.12			& \phantom{0}85.77\%					\\
					& 60	& 265.60			& 226.80			& \phantom{0}85.39\%					& 214.64			& \phantom{0}80.81\%					& 204.32			& \phantom{0}76.93\%					\\
					& 120	& 228.24			& 182.36			& \phantom{0}79.90\%					& 166.28			& \phantom{0}72.85\%					& 160.48			& \phantom{0}70.31\%					\\ \hline
\multirow{4}{*}{4}	& 1		& 132.48			& 131.84			& \phantom{0}\cellcolor{green!25}99.52\%& 131.76			& \phantom{0}\cellcolor{green!25}99.46\%& 131.00			& \phantom{0}\cellcolor{green!25}98.88\%\Tstrut\\ 
					& 2		& 134.12			& 133.56			& \phantom{0}\cellcolor{green!25}99.58\%& 132.00			& \phantom{0}\cellcolor{green!25}98.42\%& 132.00			& \phantom{0}\cellcolor{green!25}98.42\%\\ 
					& 4		& 132.28			& 131.00			& \phantom{0}\cellcolor{green!25}99.03\%& 130.92			& \phantom{0}\cellcolor{green!25}98.97\%& 130.28			& \phantom{0}\cellcolor{green!25}98.49\%\\
					& 7		& 129.04			& 129.44			& \cellcolor{green!25}100.31\%			& 129.40			& \cellcolor{green!25}100.28\%			& 127.36			& \phantom{0}\cellcolor{green!25}98.70\%\\
					& 14	& 127.08			& 126.68			& \phantom{0}\cellcolor{green!25}99.69\%& 124.36			& \phantom{0}\cellcolor{green!25}97.86\%& 121.36			& \phantom{0}\cellcolor{green!25}95.50\%\\
					& 30	& 121.12			& 115.96			& \phantom{0}\cellcolor{green!25}95.74\%& 112.40			& \phantom{0}92.80\%					& 111.68			& \phantom{0}92.21\%					\\
					& 60	& 111.32			& 103.76			& \phantom{0}93.21\%					& \phantom{0}97.88	& \phantom{0}87.93\%					& \phantom{0}96.80	& \phantom{0}86.96\%					\\ 
					& 120	& \phantom{0}96.00	& \phantom{0}81.84	& \phantom{0}85.25\%					& \phantom{0}77.72	& \phantom{0}80.96\%					& \phantom{0}75.68	& \phantom{0}78.83\%					\\ \hline
\multirow{8}{*}{7}	& 1		& \phantom{0}61.16	& \phantom{0}62.40	& \cellcolor{green!25}102.03\%			& \phantom{0}62.00	& \cellcolor{green!25}101.37\%			& \phantom{0}60.80	& \phantom{0}\cellcolor{green!25}99.41\%\Tstrut\\ 
					& 2		& \phantom{0}62.52	& \phantom{0}62.88	& \cellcolor{green!25}100.58\%			& \phantom{0}60.68	& \phantom{0}\cellcolor{green!25}97.06\%& \phantom{0}61.44	& \phantom{0}\cellcolor{green!25}98.27\%\\ 
					& 4		& \phantom{0}61.64	& \phantom{0}60.32	& \phantom{0}\cellcolor{green!25}97.86\%& \phantom{0}61.36	& \phantom{0}\cellcolor{green!25}99.55\%& \phantom{0}61.24	& \phantom{0}\cellcolor{green!25}99.35\%\\
					& 7		& \phantom{0}62.28	& \phantom{0}61.44	& \phantom{0}\cellcolor{green!25}98.65\%& \phantom{0}60.40	& \phantom{0}\cellcolor{green!25}96.98\%& \phantom{0}60.52	& \phantom{0}\cellcolor{green!25}97.17\%\\
					& 14	& \phantom{0}60.48	& \phantom{0}60.48	& \cellcolor{green!25}100.00\%			& \phantom{0}58.76	& \phantom{0}\cellcolor{green!25}97.16\%& \phantom{0}58.56	& \phantom{0}\cellcolor{green!25}96.83\%\\
					& 30	& \phantom{0}55.72	& \phantom{0}56.24	& \cellcolor{green!25}100.93\%			& \phantom{0}55.12	& \phantom{0}\cellcolor{green!25}98.92\%& \phantom{0}52.52	& \phantom{0}94.26\%					\\
					& 60	& \phantom{0}51.80	& \phantom{0}51.88	& \cellcolor{green!25}100.15\%			& \phantom{0}48.32	& \phantom{0}93.28\%					& \phantom{0}47.12	& \phantom{0}90.97\%					\\
					& 120	& \phantom{0}44.44	& \phantom{0}44.20	& \phantom{0}\cellcolor{green!25}99.46\%& \phantom{0}40.00	& \phantom{0}90.01\%					& \phantom{0}38.60	& \phantom{0}86.86\%					\\ \hline
\multirow{4}{*}{14}	& 1		& \phantom{0}22.40	& \phantom{0}22.68	& \cellcolor{green!25}101.25\%			& \phantom{0}22.72	& \cellcolor{green!25}101.43\%			& \phantom{0}22.28	& \phantom{0}\cellcolor{green!25}99.46\%\Tstrut\\ 
					& 2		& \phantom{0}21.72	& \phantom{0}21.88	& \cellcolor{green!25}100.74\%			& \phantom{0}22.32	& \cellcolor{green!25}102.76\%			& \phantom{0}21.96	& \cellcolor{green!25}101.10\%			\\ 
					& 4		& \phantom{0}22.96	& \phantom{0}23.24	& \cellcolor{green!25}101.22\%			& \phantom{0}23.52	& \cellcolor{green!25}102.44\%			& \phantom{0}23.28	& \cellcolor{green!25}101.39\%			\\
					& 7		& \phantom{0}23.16	& \phantom{0}23.20	& \cellcolor{green!25}100.17\%			& \phantom{0}23.20	& \cellcolor{green!25}100.17\%			& \phantom{0}23.20	& \cellcolor{green!25}100.17\%			\\
					& 14	& \phantom{0}22.36	& \phantom{0}22.12	& \phantom{0}\cellcolor{green!25}98.93\%& \phantom{0}22.12	& \phantom{0}\cellcolor{green!25}98.93\%& \phantom{0}22.12	& \phantom{0}\cellcolor{green!25}98.93\%\\
					& 30	& \phantom{0}19.84	& \phantom{0}19.84	& \cellcolor{green!25}100.00\%			& \phantom{0}19.80	& \phantom{0}\cellcolor{green!25}99.80\%& \phantom{0}20.00	& \cellcolor{green!25}100.81\%			\\
					& 60	& \phantom{0}19.16	& \phantom{0}19.16	& \cellcolor{green!25}100.00\%			& \phantom{0}19.04	& \phantom{0}\cellcolor{green!25}99.37\%& \phantom{0}19.04	& \phantom{0}\cellcolor{green!25}99.37\%\\
					& 120	& \phantom{0}16.40	& \phantom{0}16.40	& \cellcolor{green!25}100.00\%			& \phantom{0}16.12	& \phantom{0}\cellcolor{green!25}98.29\%& \phantom{0}16.36	& \phantom{0}\cellcolor{green!25}99.76\%\\ 
\end{tabular}

\end{table*}

\begin{table*}[hbt!]
	\caption{Comparison between the number of transplants measured for the conventional approach and the privacy-preserving approach (for different values for the latency $ L $) with a departure rate of $ 1/800 $ and a match refusal probability of $ 20\% $. The percentages state how close the privacy-preserving approach comes to the conventional approach.\label{tab:simulation_comp_dep800}}
	\small
\centering
\begin{tabular}{c | c | c | c | c | c | c | c | c}
\multicolumn{1}{c}{}& \multicolumn{1}{c}{} & \multicolumn{1}{c}{\underline{conv.\ model}}	& \multicolumn{6}{c}{\underline{privacy-preserving model}} 	\\
arrival				& match run		& \multirow{2}{*}{transplants}				& \multicolumn{2}{c}{$ L = 1\textit{ms} $} & \multicolumn{2}{c}{$ L = 10\textit{ms} $} & \multicolumn{2}{c}{$ L = 20\textit{ms} $} \Tstrut\\
rate				& interval		&  					& transplants		& percentage		& transplants 		& percentage		& transplants		& percentage\\ \hline
\multirow{4}{*}{1}	& 1				& 751.52			& 740.52			& \phantom{0}\cellcolor{green!25}98.54\%& 731.80			& \phantom{0}\cellcolor{green!25}97.38\%& 730.80			& \phantom{0}\cellcolor{green!25}97.24\%	\Tstrut\\ 
					& 2				& 748.60			& 732.80			& \phantom{0}\cellcolor{green!25}97.89\%& 725.60			& \phantom{0}\cellcolor{green!25}96.93\%& 718.88			& \phantom{0}\cellcolor{green!25}96.03\%	\\ 
					& 4				& 747.80			& 717.68			& \phantom{0}\cellcolor{green!25}95.97\%& 707.88			& \phantom{0}94.66\%					& 703.44			& \phantom{0}94.07\%						\\
					& 7				& 744.36			& 701.28			& \phantom{0}94.21\%					& 689.32			& \phantom{0}92.61\%					& 679.16			& \phantom{0}91.24\%						\\
					& 14			& 730.48			& 668.00			& \phantom{0}91.45\%					& 644.56			& \phantom{0}88.24\%					& 636.24			& \phantom{0}87.10\%						\\
					& 30			& 696.48			& 601.48			& \phantom{0}86.36\%					& 568.36			& \phantom{0}81.60\%					& 561.64			& \phantom{0}80.64\%						\\
					& 60			& 659.52			& 528.28			& \phantom{0}80.10\%					& 486.92			& \phantom{0}73.83\%					& 472.80			& \phantom{0}71.69\%						\\ 
					& 120			& 589.96			& 424.52			& \phantom{0}71.96\%					& 381.04			& \phantom{0}64.59\%					& 369.60			& \phantom{0}62.65\%						\\ \hline
\multirow{8}{*}{2}	& 1				& 332.72			& 329.52			& \phantom{0}\cellcolor{green!25}99.04\%& 327.64			& \phantom{0}\cellcolor{green!25}98.47\%& 325.36			& \phantom{0}\cellcolor{green!25}97.79\%	\Tstrut\\ 
					& 2				& 331.08			& 325.52			& \phantom{0}\cellcolor{green!25}98.32\%& 327.84			& \phantom{0}\cellcolor{green!25}99.02\%& 323.92			& \phantom{0}\cellcolor{green!25}97.84\%	\\ 
					& 4				& 331.32			& 322.36			& \phantom{0}\cellcolor{green!25}97.30\%& 321.52			& \phantom{0}\cellcolor{green!25}97.04\%& 321.12			& \phantom{0}\cellcolor{green!25}96.92\%	\\
					& 7				& 328.44			& 317.24			& \phantom{0}\cellcolor{green!25}96.59\%& 313.20			& \phantom{0}\cellcolor{green!25}95.36\%& 310.88			& \phantom{0}94.65\%						\\
					& 14			& 321.76			& 306.96			& \phantom{0}\cellcolor{green!25}95.40\%& 299.52			& \phantom{0}93.09\%					& 296.16			& \phantom{0}92.04\%						\\
					& 30			& 310.68			& 282.40			& \phantom{0}90.90\%					& 268.96			& \phantom{0}86.57\%					& 262.88			& \phantom{0}84.61\%						\\
					& 60			& 291.20			& 250.20			& \phantom{0}85.92\%					& 233.80			& \phantom{0}80.29\%					& 225.64			& \phantom{0}77.49\%						\\
					& 120			& 260.20			& 206.16			& \phantom{0}79.23\%					& 185.88			& \phantom{0}71.44\%					& 180.40			& \phantom{0}69.33\%						\\ \hline
\multirow{4}{*}{4}	& 1				& 141.24			& 141.64			& \cellcolor{green!25}100.28\%			& 141.32			& \cellcolor{green!25}100.06\%			& 141.16			& \phantom{0}\cellcolor{green!25}99.94\%	\Tstrut\\ 
					& 2				& 141.76			& 140.92			& \phantom{0}\cellcolor{green!25}99.41\%& 141.76			& \cellcolor{green!25}100.00\%			& 140.44			& \phantom{0}\cellcolor{green!25}99.07\%	\\ 
					& 4				& 141.52			& 137.72			& \phantom{0}\cellcolor{green!25}97.31\%& 139.32			& \phantom{0}\cellcolor{green!25}98.45\%& 140.04			& \phantom{0}\cellcolor{green!25}98.95\%	\\
					& 7				& 141.60			& 138.24			& \phantom{0}\cellcolor{green!25}97.63\%& 136.84			& \phantom{0}\cellcolor{green!25}96.64\%& 136.64			& \phantom{0}\cellcolor{green!25}96.50\%	\\
					& 14			& 137.60			& 134.40			& \phantom{0}\cellcolor{green!25}97.67\%& 133.76			& \phantom{0}\cellcolor{green!25}97.21\%& 132.92			& \phantom{0}\cellcolor{green!25}96.60\%	\\
					& 30			& 130.56			& 126.24			& \phantom{0}\cellcolor{green!25}96.69\%& 121.48			& \phantom{0}93.05\%					& 117.84			& \phantom{0}90.26\%						\\
					& 60			& 123.36			& 115.68			& \phantom{0}93.77\%					& 107.56			& \phantom{0}87.19\%					& 106.40			& \phantom{0}86.25\%						\\ 
					& 120			& 111.12			& \phantom{0}97.40	& \phantom{0}87.65\%					& \phantom{0}88.28	& \phantom{0}79.45\%					& \phantom{0}84.76	& \phantom{0}76.28\%						\\ \hline
\multirow{8}{*}{7}	& 1				& \phantom{0}67.28	& \phantom{0}67.32	& \cellcolor{green!25}100.06\%			& \phantom{0}66.24	& \phantom{0}\cellcolor{green!25}98.45\%& \phantom{0}68.24	& \cellcolor{green!25}101.43\%				\Tstrut\\ 
					& 2				& \phantom{0}67.92	& \phantom{0}66.28	& \phantom{0}\cellcolor{green!25}97.59\%& \phantom{0}68.00	& \cellcolor{green!25}100.12\%			& \phantom{0}66.40	& \phantom{0}\cellcolor{green!25}97.76\%	\\ 
					& 4				& \phantom{0}67.24	& \phantom{0}66.60	& \phantom{0}\cellcolor{green!25}99.05\%& \phantom{0}65.72	& \phantom{0}\cellcolor{green!25}97.74\%& \phantom{0}66.76	& \phantom{0}\cellcolor{green!25}99.29\%	\\
					& 7				& \phantom{0}66.76	& \phantom{0}66.24	& \phantom{0}\cellcolor{green!25}99.22\%& \phantom{0}66.48	& \phantom{0}\cellcolor{green!25}99.58\%& \phantom{0}65.72	& \phantom{0}\cellcolor{green!25}98.44\%	\\
					& 14			& \phantom{0}65.36	& \phantom{0}64.84	& \phantom{0}\cellcolor{green!25}99.20\%& \phantom{0}63.96	& \phantom{0}\cellcolor{green!25}97.86\%& \phantom{0}63.68	& \phantom{0}\cellcolor{green!25}97.43\%	\\
					& 30			& \phantom{0}62.08	& \phantom{0}61.00	& \phantom{0}\cellcolor{green!25}98.26\%& \phantom{0}59.84	& \phantom{0}\cellcolor{green!25}96.39\%& \phantom{0}58.76	& \phantom{0}94.65\%						\\
					& 60			& \phantom{0}60.76	& \phantom{0}55.28	& \phantom{0}90.98\%					& \phantom{0}54.48	& \phantom{0}89.66\%					& \phantom{0}54.52	& \phantom{0}89.73\%						\\
					& 120			& \phantom{0}53.32	& \phantom{0}48.52	& \phantom{0}91.00\%					& \phantom{0}46.80	& \phantom{0}87.77\%					& \phantom{0}46.08	& \phantom{0}86.42\%						\\ \hline
\multirow{4}{*}{14}	& 1				& \phantom{0}25.05	& \phantom{0}25.64	& \cellcolor{green!25}102.40\%			& \phantom{0}25.28	& \cellcolor{green!25}100.96\%			& \phantom{0}25.44	& \cellcolor{green!25}101.60\%				\Tstrut\\ 
					& 2				& \phantom{0}26.44	& \phantom{0}25.48	& \phantom{0}\cellcolor{green!25}96.37\%& \phantom{0}24.96	& \phantom{0}94.40\%					& \phantom{0}25.24	& \phantom{0}\cellcolor{green!25}95.46\%	\\ 
					& 4				& \phantom{0}25.84	& \phantom{0}25.12	& \phantom{0}\cellcolor{green!25}97.21\%& \phantom{0}24.92	& \phantom{0}\cellcolor{green!25}96.44\%& \phantom{0}25.44	& \phantom{0}\cellcolor{green!25}98.45\%	\\
					& 7				& \phantom{0}25.08	& \phantom{0}25.40	& \cellcolor{green!25}101.28\%			& \phantom{0}25.32	& \cellcolor{green!25}100.96\%			& \phantom{0}24.88	& \phantom{0}\cellcolor{green!25}99.20\%	\\
					& 14			& \phantom{0}24.64	& \phantom{0}24.72	& \cellcolor{green!25}100.32\%			& \phantom{0}24.40	& \phantom{0}\cellcolor{green!25}99.03\%& \phantom{0}23.96	& \phantom{0}\cellcolor{green!25}97.24\%	\\
					& 30			& \phantom{0}23.44	& \phantom{0}23.40	& \phantom{0}\cellcolor{green!25}99.83\%& \phantom{0}23.92	& \cellcolor{green!25}102.05\%			& \phantom{0}23.08	& \phantom{0}\cellcolor{green!25}98.46\%	\\
					& 60			& \phantom{0}22.32	& \phantom{0}22.32	& \cellcolor{green!25}100.00\%			& \phantom{0}21.48	& \phantom{0}\cellcolor{green!25}96.24\%& \phantom{0}21.32	& \phantom{0}\cellcolor{green!25}95.52\%	\\
					& 120			& \phantom{0}20.52	& \phantom{0}20.64	& \cellcolor{green!25}100.58\%			& \phantom{0}20.08	& \phantom{0}\cellcolor{green!25}97.86\%& \phantom{0}17.96	& \phantom{0}87.52\%						\\ 
\end{tabular}

\end{table*}

\section{Average Waiting Time}\label{app:waiting_time}

\begin{table*}[hbt!]
	\caption{Comparison between the average waiting time measured for the conventional approach and the privacy-preserving approach (for different values for the latency $ L $) with a departure rate of $ 1/400 $ and a match refusal probability of $ 20\% $. The percentages state how much larger or smaller the average waiting time is in the privacy-preserving approach compared to the conventional approach.\label{tab:simulation_comp_awt02}}
	\small
\centering
\begin{tabular}{c | c | c | c | c | c | c | c | c}
	\multicolumn{1}{c}{}& \multicolumn{1}{c}{} & \multicolumn{1}{c}{\underline{conv.\ model}}	& \multicolumn{6}{c}{\underline{privacy-preserving model}} 	\\
arrival				& match run		& avgerage waiting				& \multicolumn{2}{c}{$ L = 1\textit{ms} $} & \multicolumn{2}{c}{$ L = 10\textit{ms} $} & \multicolumn{2}{c}{$ L = 20\textit{ms} $} \Tstrut\\
	rate				& interval	& time [days]	& time [days]		& percentage							& time [days]& percentage				& time [days]& percentage\\ \hline
	\multirow{4}{*}{1}	& 1		& \phantom{0}91.44	& \phantom{0}96.58	& \phantom{0}5.32\%						& \phantom{0}96.71	& \phantom{0}5.45\%						& \phantom{0}98.09	& \phantom{0}6.78\%						\Tstrut\\ 
						& 2		& \phantom{0}92.59	& \phantom{0}99.95	& \phantom{0}7.36\%						& 106.26			& 12.86\%								& 102.28			& \phantom{0}9.47\%						\\ 
						& 4		& \phantom{0}93.01	& 103.46			& 10.10\%								& 109.91			& 15.38\%								& 108.39			& 14.19\%								\\
						& 7		& 96.54				& 109.30			& 11.67\%								& 114.81			& 15.91\%								& 118.48			& 18.52\%								\\
						& 14	& 100.80			& 119.41			& 15.58\%								& 125.72			& 19.82\%								& 126.44			& 20.28\%								\\
						& 30	& 115.77			& 141.73			& 18.32\%								& 149.08			& 22.34\%								& 152.15			& 23.91\%								\\
						& 60	& 140.45			& 170.42			& 17.59\%								& 178.00			& 21.10\%								& 184.44			& 23.85\%								\\ 
						& 120	& 179.36			& 210.67			& 14.86\%								& 219.37			& 18.24\%								& 221.60			& 19.06\%								\\ \hline
	\multirow{8}{*}{2}	& 1		& 103.30			& 103.22			& \cellcolor{green!25}-0.08\%			& 104.40			& \phantom{0}\cellcolor{green!25}1.05\%	& 105.99			& \phantom{0}\cellcolor{green!25}2.54\%	\Tstrut\\ 
						& 2		& 100.79			& 104.32			& \phantom{0}\cellcolor{green!25}3.38\%	& 109.57			& \phantom{0}8.01\%						& 106.13			& \phantom{0}5.03\%						\\ 
						& 4		& 101.62			& 107.51			& \phantom{0}5.48\%						& 112.00			& \phantom{0}9.27\%						& 112.90			& \phantom{0}9.99\%						\\
						& 7		& 107.50			& 114.67			& \phantom{0}6.25\%						& 118.37			& \phantom{0}9.18\%						& 121.80			& 11.74\%								\\
						& 14	& 109.13			& 122.67			& 11.04\%								& 124.03			& 12.01\%								& 127.88			& 14.66\%								\\
						& 30	& 123.85			& 140.78			& 12.03\%								& 145.62			& 14.95\%								& 150.03			& 17.45\%								\\
						& 60	& 146.13			& 166.34			& 12.15\%								& 175.29			& 16.64\%								& 179.49			& 18.59\%								\\
						& 120	& 185.77			& 201.72			& \phantom{0}7.91\%						& 213.80			& 13.11\%								& 215.51			& 13.80\%								\\ \hline
	\multirow{4}{*}{4}	& 1		& 109.05			& 114.75			& \phantom{0}\cellcolor{green!25}4.97\%	& 116.75			& \phantom{0}6.60\%						& 114.95			& \phantom{0}5.13\%						\Tstrut\\ 
						& 2		& 114.00			& 112.60			& \cellcolor{green!25}-1.24\%			& 117.12			& \phantom{0}\cellcolor{green!25}2.66\%	& 117.87			& \phantom{0}\cellcolor{green!25}3.28\%	\\ 
						& 4		& 114.48			& 115.27			& \phantom{0}\cellcolor{green!25}0.69\%	& 120.43			& \phantom{0}\cellcolor{green!25}4.94\%	& 117.20			& \phantom{0}\cellcolor{green!25}2.32\%	\\
						& 7		& 116.98			& 120.99			& \phantom{0}3.31\%						& 126.17			& \phantom{0}7.28\%						& 124.88			& \phantom{0}6.33\%						\\
						& 14	& 118.93			& 128.03			& \phantom{0}7.11\%						& 131.83			& \phantom{0}9.79\%						& 135.44			& 12.19\%								\\
						& 30	& 134.74			& 139.74			& \phantom{0}\cellcolor{green!25}3.58\%	& 149.55			& \phantom{0}9.90\%						& 147.16			& \phantom{0}8.44\%						\\
						& 60	& 157.52			& 165.99			& \phantom{0}5.10\%						& 173.74			& \phantom{0}9.34\%						& 176.69			& 10.85\%								\\ 
						& 120	& 191.35			& 206.88			& \phantom{0}7.51\%						& 207.51			& \phantom{0}7.79\%						& 215.72			& 11.30\%								\\ \hline
	\multirow{8}{*}{7}	& 1		& 126.31			& 124.68			& \cellcolor{green!25}-1.31\%			& 121.56			& \cellcolor{green!25}-3.91\%			& 118.85			& \cellcolor{green!25}-6.28\%			\Tstrut\\ 
						& 2		& 121.08			& 125.35			& \phantom{0}\cellcolor{green!25}3.41\%	& 126.16			& \phantom{0}\cellcolor{green!25}4.03\%	& 121.05			& \cellcolor{green!25}-0.02\%			\\ 
						& 4		& 122.89			& 124.17			& \phantom{0}\cellcolor{green!25}1.03\%	& 124.14			& \phantom{0}\cellcolor{green!25}1.01\%	& 124.62			& \phantom{0}\cellcolor{green!25}1.39\%	\\
						& 7		& 119.94			& 125.92			& \phantom{0}\cellcolor{green!25}4.75\%	& 129.53			& \phantom{0}7.40\%						& 129.06			& \phantom{0}7.07\%						\\
						& 14	& 129.34			& 130.58			& \phantom{0}\cellcolor{green!25}0.95\%	& 129.76			& \phantom{0}\cellcolor{green!25}0.32\%	& 133.49			& \phantom{0}\cellcolor{green!25}3.11\%	\\
						& 30	& 140.83			& 140.46			& \cellcolor{green!25}-0.26\%			& 153.78			& \phantom{0}8.42\%						& 150.09			& \phantom{0}6.17\%						\\
						& 60	& 162.04			& 167.49			& \phantom{0}\cellcolor{green!25}3.25\%	& 173.66			& \phantom{0}6.69\%						& 169.47			& \phantom{0}\cellcolor{green!25}4.38\%	\\
						& 120	& 194.92			& 198.94			& \phantom{0}\cellcolor{green!25}2.02\%	& 210.70			& \phantom{0}7.49\%						& 212.29			& \phantom{0}8.18\%						\\ \hline
	\multirow{4}{*}{14}	& 1		& 130.20			& 132.61			& \phantom{0}\cellcolor{green!25}1.82\%	& 126.12			& \cellcolor{green!25}-3.24\%			& 128.69			& \cellcolor{green!25}-1.17\%			\Tstrut\\ 
						& 2		& 130.57			& 137.74			& \phantom{0}5.21\%						& 142.27			& \phantom{0}8.22\%						& 144.55			& \phantom{0}9.67\%						\\ 
						& 4		& 124.63			& 126.25			& \phantom{0}\cellcolor{green!25}1.28\%	& 127.13			& \phantom{0}\cellcolor{green!25}1.97\%	& 134.88			& \phantom{0}7.60\%						\\
						& 7		& 129.19			& 123.97			& \cellcolor{green!25}-4.21\%			& 124.42			& \cellcolor{green!25}-3.83\%			& 125.02			& \cellcolor{green!25}-3.34\%			\\
						& 14	& 139.12			& 139.05			& \cellcolor{green!25}-0.05\%			& 139.52			& \phantom{0}\cellcolor{green!25}0.29\%	& 138.84			& \cellcolor{green!25}-0.20\%			\\
						& 30	& 144.24			& 144.35			& \phantom{0}\cellcolor{green!25}0.08\%	& 144.44			& \phantom{0}\cellcolor{green!25}0.14\%	& 147.28			& \phantom{0}\cellcolor{green!25}2.06\%	\\
						& 60	& 163.31			& 163.43			& \phantom{0}\cellcolor{green!25}0.07\%	& 166.79			& \phantom{0}\cellcolor{green!25}2.09\%	& 167.62			& \phantom{0}\cellcolor{green!25}2.57\%	\\
						& 120	& 212.94			& 213.11			& \phantom{0}\cellcolor{green!25}0.08\%	& 212.02			& \cellcolor{green!25}-0.43\%			& 212.48			& \cellcolor{green!25}-0.22\%			\\ 
\end{tabular}
\end{table*}

Until now, we have focused on the number of transplants as the key performance indicator for kidney exchange. However, the different approaches can also be compared w.r.t.\ the average waiting time which indicates how long it takes on average for a patient to obtain a transplant.

Table~\ref{tab:simulation_comp_awt02} shows the average waiting times for the conventional approach and the privacy-preserving approach for different latencies and for departure rate $ 1/400 $ and match refusal probability $ 20\% $. Thus, it contains the average waiting times corresponding to the number of transplants given in Table~\ref{tab:simulation_comp_ref02}. The percentages for the privacy-preserving approach indicate the increase (or decrease) of the average waiting time compared to the conventional approach (e.g., the entry $ 5.32\% $ in the first row for latency $ L = 1\textit{ms} $ indicates that the runtime for this parameter constellation is $ 5.34\% $ larger than for the same parameter constellation in the conventional approach). Those entries for which the average waiting time is less than $ 5\% $ worse than in the conventional approach are highlighted in the table.

We observe that in general the average waiting time is larger for the privacy-preserving approach. We attribute this mainly to two facts. First, on average there are more transplants for the conventional approach than for the privacy-preserving approach. This leads to smaller average waiting times as in general a high number of transplants also means that transplants occur more frequently and thus the patients wait less time on average until receiving a transplant. Second, the protocol execution in the privacy-preserving approach may take up to $ 7 $ days (depending on the number of patient-donor pairs in the pool) whereas the computation of the matching algorithm in the conventional approach completes instantly in our simulation. 

However, the results also show that the average waiting time is nearly identical in both approaches for those parameter constellations where the number of transplants in the privacy-preserving approach is very close to the number of transplants in the conventional approach, i.e., for large arrival rates and small match run intervals (cf.\ Section~\ref{sub:eval}). Thus, the results for the average waiting time confirm our observation for the number of transplants that the performance impact of the privacy-preserving approach is small for those parameter values which are most likely to occur in practice.

\end{document}